\documentclass[reprint,
superscriptaddress,
 amsmath,amssymb,
 aps,
pra,
]{revtex4-2}

\usepackage[english]{babel}
\usepackage{graphicx}
\usepackage{dcolumn}
\usepackage{bm}

\usepackage{amsmath, braket} 
\usepackage{amssymb}
\usepackage{xcolor}
\usepackage{bbold}
\usepackage{hyperref}
\usepackage{comment}
\usepackage{tikz}  
\usetikzlibrary{arrows,shapes,positioning,shadows,svg.path,backgrounds,fit}

\newcommand{\op}[1]{\hat{#1}}

\begin{document}

\title{Local coherence by thermalized intra-system coupling}

\author{Michal Kolář}
\email{kolar@optics.upol.cz}
\affiliation{Department of Optics, Palack\'y University, 17. listopadu 12, 771 46 Olomouc, Czech Republic}
\author{Radim Filip}
\email{filip@optics.upol.cz}
\affiliation{Department of Optics, Palack\'y University, 17. listopadu 12, 771 46 Olomouc, Czech Republic}

\begin{abstract}
Quantum superposition of energy eigenstates can appear autonomously in a single quantum two-level system coupled to a low-temperature thermal bath, if such coupling has a proper composite nature. We propose here a principally different and more feasible approach employing engineered interactions between two-level systems being thermalized into a global Gibbs state by weakly coupled thermal bath at temperature $T$. Therefore, in such case quantum coherence appears by a different mechanism, whereas the system-bath coupling does not have to be engineered.  We demonstrate such autonomous coherence generation reaching maximum values of coherence. Moreover, it can be alternatively built up by using weaker but collective interaction with several two-level systems. This approach surpasses the coherence generated by the engineered system-bath coupling for comparable interaction strengths and directly reduces phase estimation error in quantum sensing. This represents a necessary step towards the autonomous quantum sensing.  
\end{abstract}

\maketitle

\section{Introduction}
Quantum coherence is responsible for many fundamental effects and is the essential resource of upcoming quantum technology. It is, therefore, the subject of intensive basic research that culminated from the mathematical side in the formulations of resource theory of coherence~\cite{StreltsovRevModPhys2017,ChitambarRevModPhys2019} and experimentally, by coherence manipulations with diverse systems~\cite{RegulaPRL2018,adessoPRL2018,WuPRL2018,HofheinzNat2009,Gumberidze2019,Starek2021}. 
The potential role and impact of quantum coherence is known to be extremely wide, ranging from quantum biology~\cite{huelga2013vibrations,CaoScience2020}, to quantum computing~\cite{arute2019quantum,SupremacyReview,boixo2018characterizing,Solfanelli}, quantum gravity~\cite{HostenPhysRevResearch2022} and quantum thermodynamics~\cite{Lostaglio2015,Narasimhachar2015,StreltsovRevModPhys2017,KlatzowPRL2019}.
This highly sought-for local genuine quantum feature, however, is generally known to be extremely fragile whenever a thermal bath with temperature $T$ is coupled to the system of interest. This is due to the fact that (although a formal proof still remains an open problem in general~\cite{TrushechkinQuSci2022}) it has been mathematically shown that a system coupled to a bath \textit{thermalizes} to a global Gibbs state (with the Boltzmann constant $k_B=1$ from now on) $\op{\tau}_{SB} \equiv Z_{SB}^{-1} \exp\left[-\op{H}_{SB}/T\right]$, $Z_{SB} = \mathrm{Tr}\left(\exp\left[-\op{H}_{SB}/T\right]\right)$, provided (i) the full system-bath Hamiltonian has a continuous energy spectrum and that (ii) the latter does not have either degenerate eigenvalues and Bohr frequencies~\cite{Linden2009,Farrelly2017}.
Thermal states, however, clearly do not possess any coherence with respect to the system's energy eigenbasis, thus categorizing them as \textit{passive} states in the sense that no work could be extracted from them by means of unitary operations~\cite{UzdinPRX2018}. Such considerations led researchers to treat thermal baths as a hindrance to combat in order to preserve quantum coherence~\cite{breuer2002theory}. This viewpoint, however, was recently challenged by a more careful analysis showing that, under the above mentioned assumptions, the reduced system does not thermalize but \textit{equilibrates} towards the so-called mean-force Gibbs state~\cite{Hanggi, TrushechkinQuSci2022}, i.e. 
\begin{equation}\label{mfG}
    \op{\tau}_{MF}= \mathrm{Tr}_B\left(\op{\tau}_{SB}\right) = \op{\tau} + \lambda^2 \op{\tau}_{MF}^{(2)} + \lambda^4 \op{\tau}_{MF}^{(4)} + \ldots ,
\end{equation}
with $\lambda$ quantifying the system-bath coupling strength.
We emphasize that this result can be demonstrated by several different methods, including geometric expansions based on Kubo identity~\cite{Kubo1985,Mori2008} and dynamical methods using a global master equation description~\cite{Walls1970,Carmichael_1973,breuer2002theory,Cattaneo_2019,TrushechkinPRA2021,ronzaniNature2018,PekolaRevModPhys2021,KonopikPRRes2022}.
The state $\op{\tau}_{MF}$ differs from the reduced Gibbs state of the system 
\begin{eqnarray}
\op{\tau} = Z^{-1}\exp\left[-\op{H}^{tot}_{S}/T\right],\,Z = \mathrm{Tr}\left(\exp\left[-\op{H}_{S}^{tot}/T\right]\right)
\label{eq:gibbs-tot}
\end{eqnarray}
outside the ultra-weak coupling limit (where the system-bath Hamiltonian is neglected), being determined solely by the system Hamiltonian $\op{H}^{tot}_{S}$. Recent research has demonstrated that properly engineered system-bath interactions could be used in order to induce steady-state coherence (with respect to $\op{H}_S^{tot}$)~\cite{giacomoPRL2018,GUARNIERIPLA2020,purkayastha2020tunable,
TrushechkinQuSci2022,AndersPRL2021strongGibbs,slobodeniuk2021extraction}. 

\begin{figure*}[ht]
\begin{tikzpicture} 
 \node (img1)  {\includegraphics[width=.9\columnwidth]{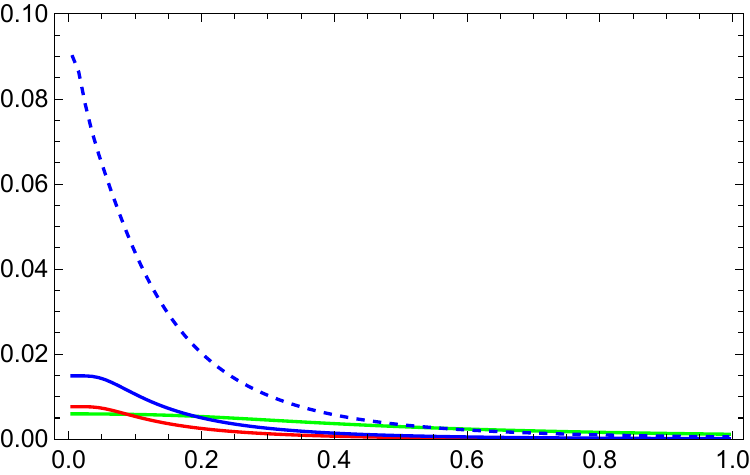}};
  \draw[red,thick,<-] (-2.79,-1.7) -- (2.1,.2);
  \draw[green,thick,<-] (-1.2,-1.8) -- (2.1,-1.2);
  \draw[blue,thick,<-] (-2.79,-1.4) -- (1.2,1.4);
  \draw[blue,dashed,thick,<-] (-2.8,.9) -- (-1.6,1.4);
\node[above=of img1, node distance=0cm, yshift=-5.8cm,xshift=2.8cm] (img2)  {\includegraphics[width=.09\linewidth]{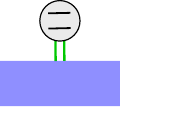}}; 
\node[above=of img1, node distance=0cm, yshift=-4.2cm,xshift=2.8cm] (img2)  {\includegraphics[width=.12\linewidth]{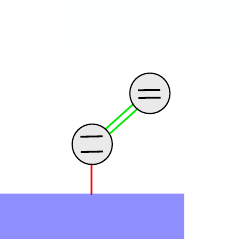}}; 
\node[above=of img1, node distance=0cm, yshift=-3cm,xshift=-1.3cm] (img2)  {\includegraphics[width=.1\linewidth]{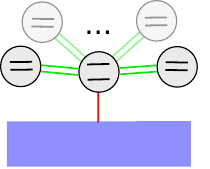}}; 
\node[above=of img1, node distance=0cm, yshift=-3cm,xshift=1.5cm] (img2)  {\includegraphics[width=.1\linewidth]{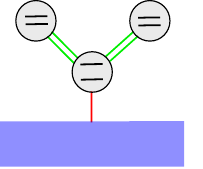}};  
  \node[above=of img1, node distance=0cm, yshift=-.8cm,xshift=.3cm] {Simplified\;scheme\;overcoming\;previous\;autonomous};
  \node[above=of img1, node distance=0cm, yshift=-1.2cm,xshift=.3cm] {coherence\;generation\;benchmark\;(green\;curve)};
  \node[above=of img1, node distance=0cm, yshift=-6.5cm,xshift=0.2cm] {$T/\omega_2$};
  \node[above=of img1, node distance=0cm, yshift=-1.9cm,xshift=3cm] {{\color{black}$\bf{(a)}$}};
    \node[above=of img1, node distance=0cm, yshift=-4.1cm,xshift=3cm] {{\color{black}$T$}};
    \node[above=of img1, node distance=0cm, yshift=-5.5cm,xshift=2.9cm] {{\color{black}$T$}};
    \node[above=of img1, node distance=0cm, yshift=-5.cm,xshift=3cm] {{\color{black}$\omega_2$}};
    \node[above=of img1, node distance=0cm, yshift=-3.5cm,xshift=3cm] {{\color{black}$\omega_2$}};
    \node[above=of img1, node distance=0cm, yshift=-2.9cm,xshift=-1.cm] {{\color{black}$T$}};
    \node[above=of img1, node distance=0cm, yshift=-1.7cm,xshift=-1.4cm] {{\color{black}$8\,{\rm x}$}};
    \node[above=of img1, node distance=0cm, yshift=-2.9cm,xshift=2.cm] {{\color{black}$T$}};
  \node[left=of img1, node distance=0cm, rotate=90, anchor=center, yshift=-.9cm,xshift=-0.cm] {{\color{black}TLS\;coherence}};
\end{tikzpicture}
\hfill
\begin{tikzpicture}
   \node (img2)  {\includegraphics[width=.9\columnwidth]{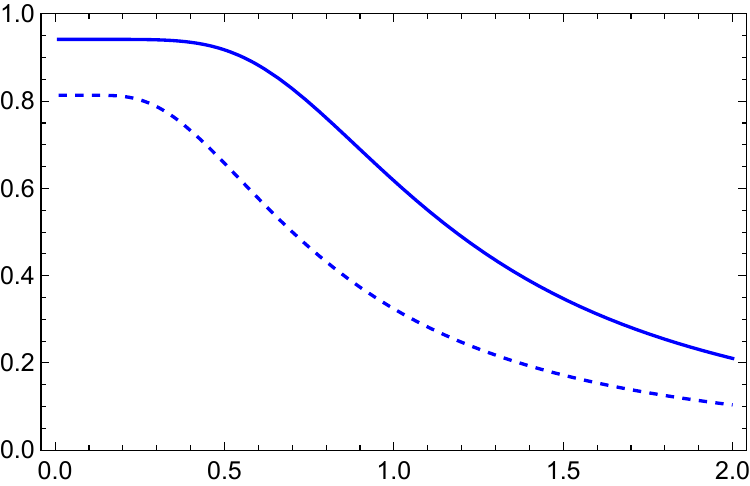}};
   \node[above=of img2, node distance=0cm, yshift=-6.7cm,xshift=0.2cm] {$T/\omega_1$};
   \draw[blue,thick,<-] (-.1,1.1) -- (1.,1.6);
   \draw[blue,dashed,thick,<-] (-1.5,.5) -- (-2.3,-.7);
   \node[above=of img2, node distance=0cm, yshift=-5.6cm,xshift=-2.9cm] {{\color{black}$T$}};
   \node[above=of img2, node distance=0cm, yshift=-2.9cm,xshift=1.8cm] {{\color{black}$T$}};
   \node[above=of img2, node distance=0cm, yshift=-1.9cm,xshift=3cm] {{\color{black}$\bf{(b)}$}};
   \node[above=of img2, node distance=0cm, yshift=-1.7cm,xshift=1.2cm] {{\color{black}$\rm{4\,x}$}};
   \node[above=of img2, node distance=0cm, yshift=-1.8cm,xshift=-.5cm] {{\color{black}$H_4\doteq 0.89$}};
   \node[above=of img2, node distance=0cm, yshift=-2.8cm,xshift=-2.7cm] {{\color{black}$H_1\doteq 0.66$}};
   \node[above=of img2, node distance=0cm, yshift=-4.9cm,xshift=-2.7cm] {{\color{black}$\omega_1$}};
   \node[above=of img2, node distance=0cm, yshift=-3cm,xshift=1.3cm] (img2)  {\includegraphics[width=.11\linewidth]{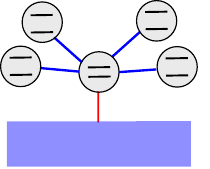}};
   \node[above=of img2, node distance=0cm, yshift=-5.5cm,xshift=-3.5cm] (img4)  {\includegraphics[width=.11\linewidth]{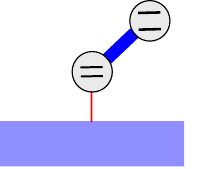}};
   \node[left=of img2, node distance=0cm, rotate=90, anchor=center, yshift=3.4cm,xshift=-1.6cm] {{\color{black}TLS\;coherence}};
   \node[above=of img2, node distance=0cm, yshift=-.8cm,xshift=-1.5cm] {Two\;methods\;to\;achieve\;significant\;autonomous};
\node[above=of img2, node distance=0cm, yshift=-1.1cm,xshift=-1.5cm] {coherence\;of\;TLS};
\end{tikzpicture}
\caption{(a) Quantitative comparison of different ways of autonomous coherence generation and their temperature dependence. TLS coupled to a low-temperature $T$ bath (green curve) via a {\it composite} coupling (parallel green lines) \cite{giacomoPRL2018} sets a benchmark for autonomous coherence generation. The used parameters are: the frequency of the relevant TLS $\omega_2=1$, spectral density cutoff $\Omega=5\,\omega_2$ and coupling parameter $\lambda=10^{-3}$, assuming Ohmic bath \cite{giacomoPRL2018}. This benchmark can be overcome (red curve) in the settings in which the relevant TLS is interacting with the bath via {\it non-composite} coupling (single red line) in the weak coupling limit {\it and simultaneously} with an ancillary TLS via a composite type of coupling, see Eq.~\eqref{eq:2nd-hamtot}. The coupling strength is comparable as in the green-curve case, by setting $\gamma\theta\sim \lambda\Omega$, $\omega_1=0.2\,\omega_2$, and $\omega_2=1$. Such coherence-generation improvement can be further up-scaled (blue and blue-dashed curves) by employing another ancillary TLSs interacting with the relevant TLS in the same way, see Eq.~\eqref{eq:2nd-hamtot-N}. (b) Different methods to achieve significant values of autonomous TLS coherence and its temperature dependence due to thermalization by weak coupling to a heat bath. The red lines have the same meaning as in panel~(a). Note the increased width of the high-coherence plateaus due to the use of increased coupling strength compared to panel (a). One can use (dashed curve) strong coupling (thick blue line) to a single ancillary TLS, according to Eq.~\eqref{eq:1st-hamtot}, with QFI $H_1$, Eq.~\eqref{eq:qfi1}, on the plateau. Alternatively (full curve), one can employ weaker collective coupling (thin blue lines) to several ancillary TLSs according to Eq.~\eqref{eq:1st-hamtot-N}. The resulting QFI on the plateau is $H_4$, Eq.~\eqref{eq-QFI-N}.  }
\label{fig:coh-bath-vs-2nd}
\end{figure*}
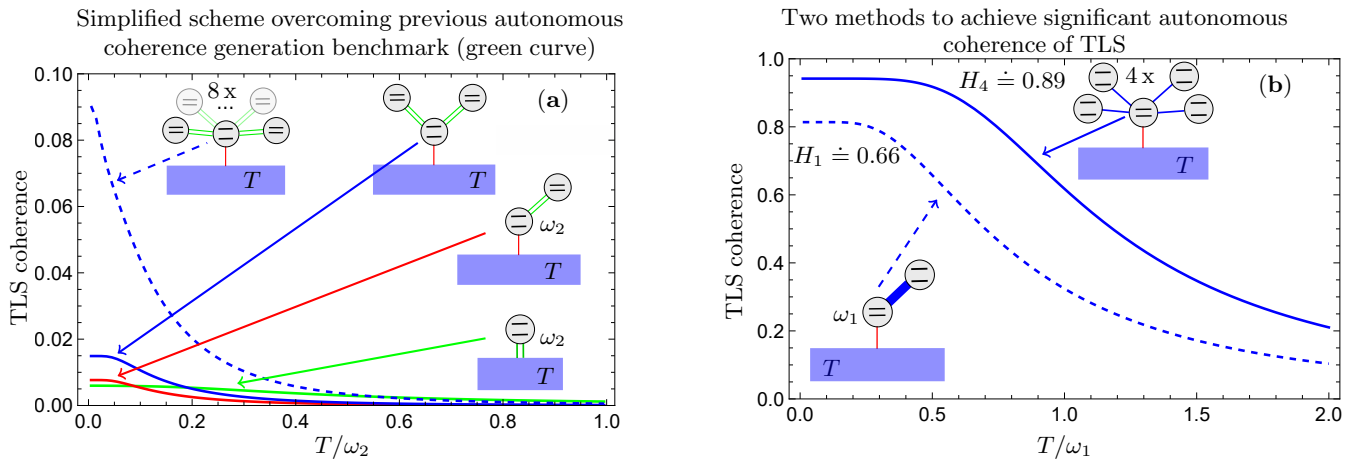

In this work we take another significant step forward along this research line but from a different angle, namely by exploiting \textit{intra-system interaction Hamiltonians}, rather than system-bath interactions, to achieve coherence in the local energy eigenbasis, see Fig.~\ref{fig:coh-bath-vs-2nd}. 
In particular, we start by considering a composite quantum system made of a {\it finite} number of several permanently mutually interacting two-level systems (TLSs). Such low-dimensional quantum systems are basic elements in many experimental platforms and key building blocks for quantum technology. We then focus on the regime of ultra-weak system-bath coupling~\cite{RicardoPRA2021}, i.e. truncate the steady-state solution in Eq.~\eqref{mfG} to the lowest order, thus assuming the system to be described by a thermal Gibbs state $\op{\tau}$, Eq.~\eqref{eq:gibbs-tot}. We will now turn our attention to the conditions on the structure of the Hamiltonian $\op{H}_S^{tot}=\sum_j\op{H}^{(j)}_S+\op{V}_{int}$ allowing for non-zero coherence in the local energy eigenbases of $\op{H}^{(j)}_S$, assuming the mutual TLSs interaction $\op{V}_{int}$.
We stress out that thermal states~\eqref{eq:gibbs-tot} can either be prepared or induced by the coupling of the system with a thermal bath in the ultra-weak coupling regime which, in the long-time limit, guarantees thermalization~\cite{TrushechkinQuSci2022}. 
Moreover, this approach, where the specific interactions are assumed between the subsystems, instead of the system and bath, have the potential to {\it quantitatively overcome} the coherence generated in the latter scheme \cite{giacomoPRL2018}, in the comparable coupling strength regime, and, in principle, generate significant values of the coherence approaching the maximum value, if stronger couplings are assumed, see Fig.~\ref{fig:coh-bath-vs-2nd}(a)-(b). This enhanced quantum coherence can be directly used to reduce estimation error in a Ramsey-type interferometer \cite{AVSBirrittella} quantified by the Quantum Fisher Information (QFI)~\cite{BraunsteinPRL1994} and is necessary to obtain autonomous quantum sensors.


\section{Local Coherence in thermal states}
Throughout this paper, we characterize the amount of local coherence in the $j$-th subsystem's state $\op{\rho}_{j}$ through the  $l_1-$norm of coherence~\cite{baumgratz2014} as $C_{l_1}(\op{\rho}_{j})\equiv C_{j}=\\|\langle\op{\sigma}_j^x+i\op{\sigma}_j^y\rangle_{\op{\tau}}|$, where $\op{\sigma}_j^{x(y,z)}$ denote the respective Pauli operators, $\op{\tau}$ is the Gibbs state \eqref{eq:gibbs-tot}, and $\langle\bullet\rangle_{\op{\tau}}\equiv {\rm Tr}_S(\bullet\,\op{\tau})$. Within this framework, we demonstrate the necessary condition implying that if $\op{H}^{tot}_{S}$ is $\mathbb{Z}_2$ symmetric~\cite{OlavPRA2003,JustinoPRA2012chainsymmetries,GUPhysRep2017} with respect to the transformation $\op{\mathbb{Z}}\equiv\bigotimes_j\op{\sigma}_j^z$, Appendix~\ref{sec:S-symmetry}, then the state $\op{\tau}$ does not support coherence in the local energy basis of the respective TLSs, i.e.
\begin{eqnarray}
[\op{\mathbb{Z}},\op{H}^{tot}_{S}]=0\quad\Rightarrow\quad \forall j\; C_j= 0.
\label{eq:necessary-symmetry}
\end{eqnarray}
Furthermore, we show that the condition~\eqref{eq:necessary-symmetry} becomes also sufficient, if the system consists of a pair of TLS, i.e. 
\begin{eqnarray}
[\op{\mathbb{Z}},\op{H}^{tot}_{S}]=0\quad\Leftrightarrow\quad \forall j\; C_j= 0.
\label{eq:necessary-sufficient-symmetry}
\end{eqnarray}
Consequently, in this work we will consider Hamiltonians involving $\op{\mathbb{Z}}$ symmetry-breaking interactions of the form of $\sigma^{x(y)}_1\sigma^z_2$, typically used to model superconducting-circuit platforms~\cite{BlaisPRA2004Transmon1,KochPRA2007Transmon2,Hamedani_Pekola_Entropy2021}.

\subsection{Directly induced coherence} 
\label{sub-direct}
\begin{figure*}[ht]
\begin{tikzpicture} 
  \node (img1)  {\includegraphics[width=.6\columnwidth]{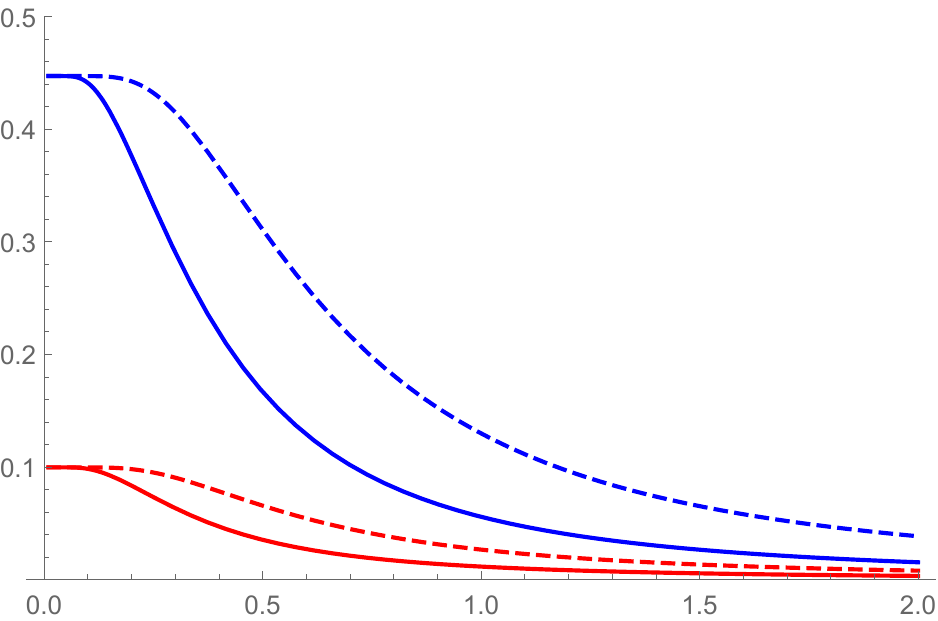}};
  \node[above=of img1, node distance=0cm, yshift=-1.2cm,xshift=.0cm] {Direct\;coherence\;generation};
  \node[above=of img1, node distance=0cm, yshift=-5cm,xshift=0cm] {$T/\omega_1$};
   \node[above=of img1, node distance=0cm, yshift=-1.8cm,xshift=1.7cm] {{\color{black}{\bf (a)}}};
   \node[above=of img1, node distance=0cm, yshift=-3cm,xshift=1cm]{{\color{red}$\gamma=0.1$}};
   \node[above=of img1, node distance=0cm, yshift=-2.5cm,xshift=1cm]{{\color{blue}$\gamma=0.5$}};
   \node[above=of img1, node distance=0cm,rotate=-60, anchor=center, yshift=-3cm,xshift=1.5cm] {{\color{blue}$\omega_2=0.5$}};
    \node[above=of img1, node distance=0cm, yshift=-2.2cm,xshift=-0.5cm] {{\color{blue}$\omega_2=1.3$}};
    \node[left=of img1, node distance=0cm, rotate=-15, anchor=center, yshift=-.2cm,xshift=2.5cm] {{\color{red}$\omega_2=1.3$}};
  \node[left=of img1, node distance=0cm, rotate=90, anchor=center, yshift=-.9cm,xshift=-0.cm] {{\color{black}$C_1$}};
\end{tikzpicture}
\hfill
\begin{tikzpicture} 
  \node (img1)  {\includegraphics[width=.6\columnwidth]{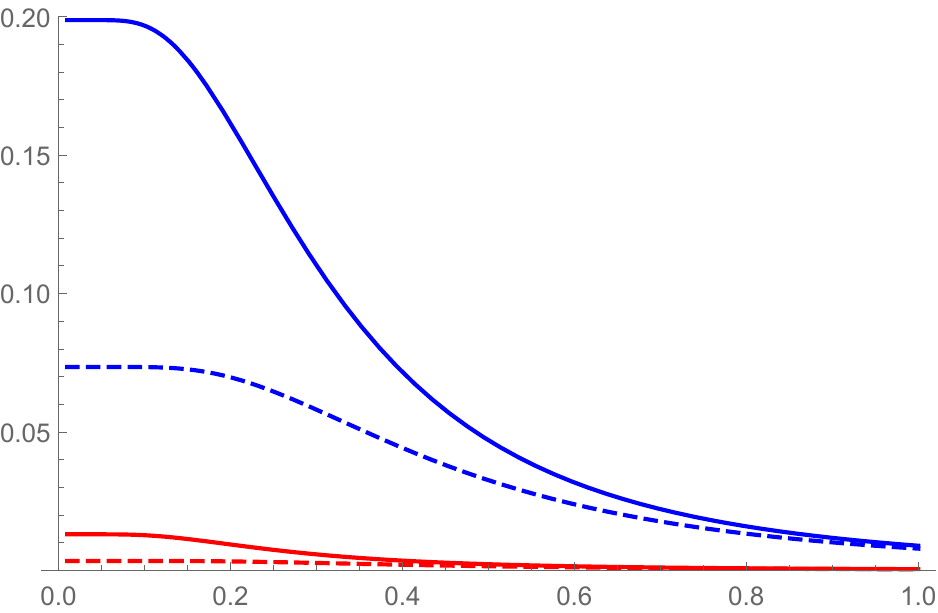}};
  \node[above=of img1, node distance=0cm, yshift=-1.2cm,xshift=.0cm] {Indirect\;coherence
  generation};
  \node[above=of img1, node distance=0cm, yshift=-1.8cm,xshift=1.7cm] {{\color{black}{\bf (b)}}};
  \node[above=of img1, node distance=0cm, yshift=-5cm,xshift=0cm] {$T/\omega_2$};
   \node[above=of img1, node distance=0cm, yshift=-3.cm,xshift=.9cm]{{\color{red}$\gamma=\theta=0.1$}};
   \node[above=of img1, node distance=0cm, yshift=-2.5cm,xshift=.9cm]{{\color{blue}$\gamma=\theta=0.5$}};
   \node[above=of img1, node distance=0cm,rotate=-50, anchor=center, yshift=-2.5cm,xshift=.5cm] {{\color{blue}$\omega_1=0.5$}};
    \node[above=of img1, node distance=0cm, yshift=-3.2cm,xshift=-1.5cm] {{\color{blue}$\omega_1=1.3$}};
    \node[left=of img1, node distance=0cm, rotate=-0, anchor=center, yshift=-1.cm,xshift=2.1cm] {{\color{red}$\omega_1=0.5$}};
    \node[left=of img1, node distance=0cm, rotate=90, anchor=center, yshift=-.9cm,xshift=-0.cm] {{\color{black}$C_2$}};
\end{tikzpicture}
\hfill
\begin{tikzpicture} 
  \node (img1)  {\includegraphics[width=.6\columnwidth]{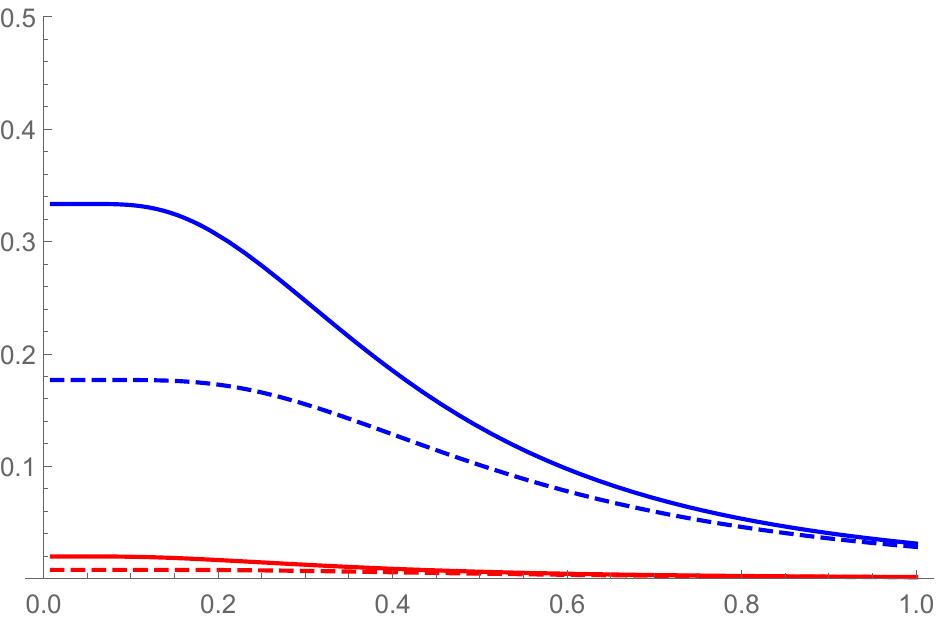}};
  \node[above=of img1, node distance=0cm, yshift=-1.3cm,xshift=.0cm] {Transferred\,direct
  generation};
  \node[above=of img1, node distance=0cm, yshift=-1.8cm,xshift=1.7cm] {{\color{black}{\bf (c)}}};
  \node[above=of img1, node distance=0cm, yshift=-5cm,xshift=0cm] {$T/\omega_0$};
   \node[above=of img1, node distance=0cm, yshift=-3cm,xshift=1.8cm]{{\color{red}$\gamma=\theta=0.1$}};
   \node[above=of img1, node distance=0cm, yshift=-2.5cm,xshift=1.8cm]{{\color{blue}$\gamma=\theta=0.5$}};
   \node[above=of img1, node distance=0cm,rotate=-0, anchor=center, yshift=-2.cm,xshift=-1.5cm] {{\color{blue}$\omega_1=0.5$}};
    \node[above=of img1, node distance=0cm, yshift=-3.2cm,xshift=-1.5cm] {{\color{blue}$\omega_1=1.3$}};
    \node[above=of img1, node distance=0cm, rotate=-0, anchor=center, yshift=-4cm,xshift=-1.5cm] {{\color{red}$\omega_1=0.5$}};
    \node[left=of img1, node distance=0cm, rotate=90, anchor=center, yshift=-.9cm,xshift=-0.cm] {{\color{black}$C_0$}};
\end{tikzpicture}
\caption{ The dependence of autonomous local coherence on $T$ for various values of parameters. (a) The direct generation  of the first TLS coherence $C_1$, Eq.~\eqref{eq:1st-hamtot}, the choice $\omega_1=1$ was made. The full red line corresponds to $\omega_2=0.5$. (b) Indirect  generation of the second TLS coherence $C_2$, Eq.~\eqref{eq:2nd-hamtot}, with the choice $\omega_2=1$. The values of corresponding parameters as well as the curves' style and colors in the respective cases were chosen the same in all panels (a-c) to underline their qualitatively different effect on the final coherence. We have set $\theta=\gamma$ for the sake of simplicity. (c) Coherence $C_0$ of the zeroth TLS, Eq.~\eqref{eq:2nd-BS-hamtot}. We set for all cases $\omega_0=1$, $\omega_2=1.3$, and red dashed line corresponds to $\omega_1=1.3$. We see qualitatively the same behavior as in (b), especially the dependence on the second TLS frequency $\omega_1$. Note considerably higher coherence values compared to the indirect generation in (b), for comparable values of parameters.  }
\label{fig:coh-1st-2nd}
\end{figure*}
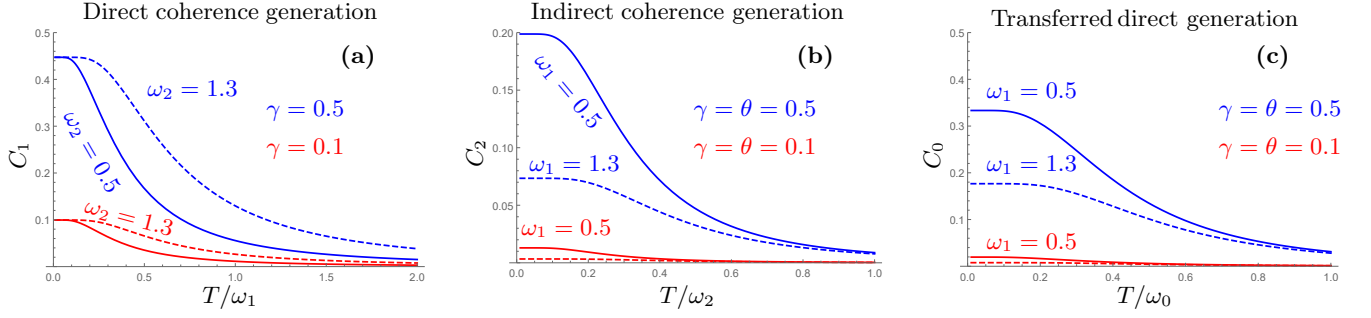

Let us begin by considering the simplest scenario involving a system made of a pair of interacting TLS, modeled by a the Hamiltonian ~\cite{BlaisPRA2004Transmon1,LisenfeldPRL2010,Hamedani_Pekola_Entropy2021,PekolaRevModPhys2021,GuthriePhysRevApp2022} (with $\hslash =1$ from now on)
\begin{eqnarray}\nonumber
\op{H}^{tot}_{S}&=& \op{H}^{(1)}_S+\op{V}^{(1,2)}_{int}+\op{H}^{(2)}_S\\ &=& \frac{\omega_1}{2}\;\op{\sigma}_1^z+\frac{\gamma}{2}\;\op{\sigma}_1^x\op{\sigma}_2^z+\frac{\omega_2}{2}\;\op{\sigma}_2^z,\quad \gamma >0.
\label{eq:1st-hamtot}
\end{eqnarray}
We stress out, that the presence of a direct external coherent drive would mean that Eq.~\eqref{eq:1st-hamtot} contains a term proportional to $\sigma_1^x$, which is {\it not} the case. Even without such external drive, the interaction has still the potential to create coherence in the first TLS by population in the second TLS, since a back-action causing dephasing on this second TLS does not play negative role. However, the final result depends on the entire Hamiltonian~\eqref{eq:1st-hamtot} and on the bath temperature.
The simplicity of the model allows for an analytical determination of the coherences, which read, see Fig.~\ref{fig:coh-1st-2nd}(a),
\begin{eqnarray}
C_1&=&\frac{\gamma\;\tanh{\left (\frac{\omega_2}{2T}\right )}\tanh{\left (\frac{\sqrt{\gamma^2+\omega_1^2}}{2T}\right )}}{\sqrt{\gamma^2+\omega_1^2}},\, C_{2}=0.
\label{eq:C1-1st}
\end{eqnarray}
The resulting coherence $C_1>0$ appears always for finite $T$, irrespective of $\omega_{1(2)},\gamma$, and  shows some typical features of thermally induced coherence of previous results~\cite{giacomoPRL2018,GUARNIERIPLA2020,AndersPRL2021strongGibbs,
TrushechkinQuSci2022,ArtaPRA2017,RicardoPRA2021}, namely monotonous decrease of $C_1$ with $T$. This implies existence of coherence maximum with respect to $T$, reached in the low-temperature regime $T\ll \gamma,\omega_{1(2)}$. Setting $\tanh x\approx [1-2\exp(-2x)]$ for $x\gg 1$ and assuming without loss of generality $\omega_{1}<\omega_{2}$, Eq.~\eqref{eq:C1-1st} yields
\begin{eqnarray}
C_1\approx \frac{\gamma}{\sqrt{\gamma^2+\omega_1^2}}\left[1-2\exp\left(-\frac{\sqrt{\gamma^2+\omega_1^2}}{T}\right)\right],
\label{eq:C1-1st-lowT}
\end{eqnarray}
showing exponentially fast convergence to the maximum value for decreasing $T$, providing an important stable "plateau" of coherence. Typical temperature-dependence profiles of high generated values of coherence are presented in Figs.~\ref{fig:coh-bath-vs-2nd}(b)~and~\ref{fig:coh-1st-2nd}(a). Equation~\eqref{eq:C1-1st-lowT} reveals interesting asymmetric role of $\omega_{1(2)}$, resulting in independence of the low-$T$ limit of $C_1$ on $\omega_{2}$. This frequency affects only the width of the plateau of exponentially fast convergence to the low-temperature limit~\eqref{eq:C1-1st-lowT}, see Fig.~\ref{fig:coh-1st-2nd}(a). Let us discuss different regimes of low-$T$ limit, Eq.~\eqref{eq:C1-1st-lowT}. If the coupling is weak $\gamma\ll\omega_1$, Eq.~\eqref{eq:C1-1st-lowT} yields $C_1\approx \gamma/\omega_1$. 

The challenging opposite regime of strong coupling $\gamma\gg\omega_1$ allows, 
in principle, to reach the maximum of coherence in the first TLS, i.e. the maximum value consistent with the assumption of thermal state~\eqref{eq:gibbs-tot} existence. In this sense, this first method of direct coherence generation is optimal, saturating inverse quadratically with $\gamma$ as $C_1\approx 1-\omega_1^2/2\gamma^2$ in the regime $T\ll\omega_1\ll\gamma$. Such maximum value of TLS coherence can be in principle reached {\it only} if the corresponding state of TLS is pure. As being the case in the low temperature limit, it suggests that the ground state of the Hamiltonian~\eqref{eq:1st-hamtot} is a product state, as can be checked  by direct diagonalization, yielding the structure $\ket{E_0}=(\cos\epsilon_0\ket{g_1}+\sin\epsilon_0\ket{e_1})\ket{g_2}$. 

Such generated coherence can be directly used, e.g. in phase estimation, where the minimal estimation error is inversely proportional to the Quantum Fisher Information $H$ (QFI)~\cite{BraunsteinPRL1994}. For example, in the low temperature limit of Eq.~\eqref{eq:C1-1st-lowT}, the QFI reads \cite{BrivioPRA2010} 
\begin{equation}
H_1=(C_1)^2=\frac{\gamma^2}{\gamma^2+\omega_1^2},
\label{eq:qfi1}
\end{equation}
showing direct positive impact of generated coherence on the feasibility of the task. The strategies how to saturate QFI $H$ are described in \cite{BrivioPRA2010}.

The high temperature limit behavior of $C_1$~\eqref{eq:C1-1st} shows asymptotically vanishing value of coherence, a general feature generally discussed in Appendix~\ref{sec:S-high-T}, presenting as well its analytic form and comparison to other coherence generating scenarios.





\subsection{Indirectly induced coherence} In cases such that the first TLS is not accessible for some reason, while the interaction between the TLSs remains as in Eq.~\eqref{eq:1st-hamtot}, the goal might be to generate coherence on the second TLS. As the interaction in~\eqref{eq:1st-hamtot} is fixed, and we assume the TLSs being {\it not} interchangeable for any reason, we analyze here the case of indirect coherence generation.  We emphasize again, that the bath is assumed to couple to the system via a standard (non-composite) interaction, representing conceptually different situation compared to~\cite{giacomoPRL2018}. As a stimulating fact, we stress out that this distinctly simplified scheme allows for quantitatively better results when comparable coupling strengths are assumed. This is graphically illustrated in Fig.~\ref{fig:coh-bath-vs-2nd}~(a). Furthermore, such scheme allows for further quantitative improvements of coherence by up-scaling the number of ancillary TLSs collectively coupled to the TLS of interest, which can be subsequently used in application. This effectively implements a build-up of intermediate-to-strong coupling effects, see Sec.~\ref{sec-N} for details.

The indirect coherence generation process is realized by the system Hamiltonian including inter-system interaction with $\gamma,\theta >0,$
\begin{eqnarray}
\op{H}^{tot}_{S}=\frac{\omega_1}{2}\;\op{\sigma}_1^z+\frac{\gamma}{2}\;\op{\sigma}_1^x\op{\sigma}_2^z+\frac{\theta}{2}\;\op{\sigma}_1^x\op{\sigma}_2^x+\frac{\omega_2}{2}\;\op{\sigma}_2^z.
\label{eq:2nd-hamtot}
\end{eqnarray}
In this regime, one can grasp the coherence-generating principle intuitively, based on a dynamical action of~\eqref{eq:2nd-hamtot} on ground states of TLSs. In such view, the coherence is directly induced on the first TLS by the population of the second TLS of interest (where it is initially {\it not} present), by the term~$\propto \gamma$, as in~\eqref{eq:1st-hamtot}. As a second-order effect, the first TLS {\it transfers} the coherence to the second TLS system of interest~\cite{slobodeniuk2021extraction}, due to the presence of interaction~$\propto~\theta$. Naturally, this indirect way is not as efficient as the direct case was.
Again, any direct external coherent drive would be accounted for by presence of a term $\op{\sigma}_{1(2)}^x$ in \eqref{eq:2nd-hamtot}, which is not the case.

Exact result for the coherence of the second TLS $C_2$ can be calculated analytically, but in this case yields too complex mathematical form, unfortunately. Therefore, we resort to approximate results and graphical representation of the numerical results, see Fig.~\ref{fig:coh-1st-2nd}(b). Similarly as in the case of directly induced coherence in previous subsection, we focus on the limit of low temperatures $T\ll \gamma,\theta,\;\omega_{1(2)}$, which yields maximum coherence $C_2$  
\begin{eqnarray}\nonumber
C_2&\approx & \frac{\theta\gamma}{{\sqrt{\omega_1^2+\gamma^2}}}\;\sqrt{\frac{ \theta^2+(\omega_2-\overline{\omega}_1)^2}{(\theta^2+\omega_2^2+\overline{\omega}_1^2)^2-4\omega_2^2\overline{\omega}_1^2}},\\ 
\nonumber
&&\overline{\omega}_1 =\sqrt{\omega_1^2+\gamma^2},\quad T\ll \gamma,\theta,\;\omega_{1(2)}, \\
C_2&\approx & \frac{\theta\gamma}{\omega_1(\omega_1+\omega_2)},\; T\ll  \gamma,\theta\ll\omega_{1(2)}.
\label{eq:2nd-lowT}
\end{eqnarray}
In this low temperature limit, both TLSs have non-vanishing coherence for any $\omega_{1(2)},\gamma$, and $\theta$. We should distinguish two regimes. In the regime of weak inter-system couplings $\theta,\gamma\ll\omega_{1(2)}$, $C_2$ is typically proportional to product of $\theta\gamma$ (last line of Eq.~\eqref{eq:2nd-lowT}). Therefore, it merely monotonically increases with these values. Recalling the result from the previous subsection for coherence $C_1\approx\gamma/\omega_1$ generated on the first TLS in the weak coupling regime, we conclude that the  coupling $\propto\theta$ further decreases the resulting coherence $C_2$ by a factor \\$\theta/(\omega_1+\omega_2)$, compared to the already small coherence $C_1$, during the backward transfer. Qualitatively, this holds even if $\theta\approx\omega_{1(2)}$.

For stronger couplings $\theta,\gamma\approx\omega_{1(2)}$, one can find for each given $\theta$ and $\omega_{1(2)}$ certain optimal value $\overline{\gamma}$ yielding maximum value of $C_2$. Such optimal value  $\overline{\gamma}$ reads
\begin{eqnarray}\nonumber
\overline{\gamma}(\theta;\omega_1,\omega_2)&\approx &\omega_1^{2/3}\left( \omega_2^{1/3}+\frac{\theta^2}{3\omega_2^{5/3}}-\frac{\theta^4}{9\omega_2^{11/3}}\right ),\\ \omega_1&\lesssim &\omega_2,\;\theta\lesssim\omega_2.
\label{eq:2nd-gamma-opt}
\end{eqnarray}
Fine tuning of the values $\theta$ and $\gamma$ according to Eq.~\eqref{eq:2nd-gamma-opt}, can yield slight increase of $C_2$ of the order of few percent. This applies strictly in the region of low $T$. By direct numerical comparison, we have found that low $T$ values of the first TLS coherence $C_1$ for the indirect generation \eqref{eq:2nd-hamtot}, are slightly lower compared to values of $C_1$ for direct generation~\eqref{eq:C1-1st-lowT}, while the difference is of the order of 1$\%$ of the actual value. By their comparison in the regime of strong couplings $\theta\approx\gamma\gg\omega_{1(2)}$ the general rule that $C_2\leq C_1$ hols as well, while $C_2/C_1\approx\theta/\sqrt{\theta^2+\gamma^2}$.

Another feature qualitatively differentiating between the direct~\eqref{eq:1st-hamtot} and indirect~\eqref{eq:2nd-hamtot} coherence generation is the structure of the respective ground states. The ground state of~\eqref{eq:2nd-hamtot} is of the type $\ket{E_0}=\alpha_0\ket{e_1e_2}+\beta_0\ket{e_1g_2}+\gamma_0\ket{g_1e_2}+\delta_0\ket{g_1g_2}$ spanning fully the basis states of a pair of TLS Hilbert space, establishing correlations between them. This has negative consequences in cases when the generated coherence is to be transferred/used in another external system. 

\subsection{Transferred directly induced coherence}  An alternative to the previous indirect method can use another (labeled by "0") TLS that can receive quantum coherence from the first TLS in  Eq.~\eqref{eq:1st-hamtot}. This can offer an option, see~\cite{slobodeniuk2021extraction}, to the method based on Hamiltonian~\eqref{eq:2nd-hamtot}, e.g. in cases when both the first and second TLSs are not accessible.   

The total Hamiltonian determining the Gibbs state $\op{\tau}$, Eq.~\eqref{eq:gibbs-tot}, with $\gamma,\theta >0$ in this case reads 
\begin{eqnarray}
\op{H}^{tot}_{S}&=&\frac{\omega_0}{2}\;\op{\sigma}_0^z+\frac{\omega_1}{2}\;\op{\sigma}_1^z+\frac{\omega_2}{2}\;\op{\sigma}_2^z\label{eq:2nd-BS-hamtot}
\\\nonumber
&+&\frac{\theta}{2}\;\op{\sigma}_0^x\op{\sigma}_1^x+\frac{\gamma}{2}\;\op{\sigma}_1^x\op{\sigma}_2^z.
\end{eqnarray}
In the low temperature limit the $C_0$ coherence is nonzero for all values of parameters, see Fig.~\ref{fig:coh-1st-2nd}(c), and has the form
\begin{eqnarray}
C_0\approx \frac{\theta\gamma}{\sqrt{(\theta\gamma)^2+\omega_0^2(\gamma^2+\omega_1^2)}},\; T\ll \gamma,\theta,\omega_{1(2)},
\label{eq:C1-2nd-BS-lowT}
\end{eqnarray}
revealing remarkable similarity to Eq.~\eqref{eq:C1-1st-lowT}. With this method, we can in principle obtain $C_0\approx 1$, if jointly achieving $\gamma\gg\omega_1$ and $\theta\gg\omega_0$ regimes, similarly as in the direct generation approach, see discussion below Eq.~\eqref{eq:C1-1st-lowT}. By substitution $\gamma\rightarrow\theta\gamma$ and $\omega_1\rightarrow\omega_0\sqrt{(\gamma^2+\omega_1^2)}$ in~\eqref{eq:C1-1st-lowT}, we can establish the parallel between these two generation methods. Therefore, we shortly outline comparison of their outcomes in the following lines. 

For low enough temperatures, the transferred coherence $C_0$ might be larger than the directly generated coherence $C_1$ from Eq.~\eqref{eq:C1-1st-lowT} taking the same values of relevant parameters $\gamma,\omega_1$ in the region $\theta/\omega_0\gtrsim 1$, assuming intermediate values of $C_1\lesssim 0.7$. In this regime, it is better to transfer the coherence to a different TLS than to use directly $C_1$ from \eqref{eq:C1-1st-lowT}. In general, the ratio of these two values is $C_0/C_1=x/\sqrt{1+C_1^2x^2}$, where $x=\theta/\omega_0$. The ratio $x$ necessary for obtaining $C_0\geq C_1$ scales as  $x\geq (1-C_1^2)^{-1/2}$. In the opposite regime where $C_1,x\ll 1$ the coherence transfer from "1" to "0" will only decrease its value, as $C_0/C_1\approx x\ll 1$.


Aside of the above comparison of the direct and transferred methods, we point out that the coherence of TLS "1" in the current model after the transfer from TLS "0" can be also enhanced by the coupling $\propto\theta$, compared to directly induced coherence $C_1$, Eq.~\eqref{eq:C1-1st-lowT}. Therefore, the very presence of TLS "0" can have a positive effect on the TLS "1" coherence, even though the coupling term $\propto\theta$ has by itself no coherence inducing potential, as it does not break the above mentioned symmetry, Eq.~\eqref{eq:necessary-symmetry}. This further underlines the non-intuitive properties of autonomously induced local coherence and its transfer, making it interesting for further theoretical and experimental investigation. 


\section{Effective strong coupling limit}
\label{sec-N}
In the following paragraphs we will show the possibility to achieve effectively the results from the above sections, valid in the strong inter-system coupling regime, even though the subsystems might be mutually coupled only weakly. Such bottom to top build-up of strong coupling effects is based on the {\it collective} interaction of the TLS of interest with finite number $N$ of the (ancillary) source TLSs, whereas all these systems may be weakly coupled to their thermal environment~\cite{ronzaniNature2018,PekolaRevModPhys2021}.
The total Hamiltonian straightforwardly generalizing the direct coherence generation method~\eqref{eq:1st-hamtot} is 
\begin{eqnarray}
\op{H}^{tot}_{S}=\frac{\omega_1}{2}\;\op{\sigma}_1^z+\sum_{j=2}^{N+1}\frac{\omega_j}{2}\;\op{\sigma}_j^z+\sum_{j=2}^{N+1}\frac{\gamma_j}{2}\;\op{\sigma}_1^x\op{\sigma}_j^z,
\label{eq:1st-hamtot-N}
\end{eqnarray}
where $\gamma_j >0,\;N\geq 1$, and we assume that all source TLSs are approximately identical, $\omega_j\approx\omega_2$, see Appendix~\ref{sec:S-enhancement} for details. In the low temperature limit $T\ll \gamma_j,\omega_{1(2)}$, we obtain for the coherence $C_1^N$, $N\geq 1$
\begin{eqnarray}
C_1^N\approx \frac{\sum_{j=2}^{N+1}\gamma_j}{\sqrt{(\sum_{j=2}^{N+1}\gamma_{j})^2+\omega_1^2}},
\label{eq:C1-1st-lowT-N}
\end{eqnarray}
in analogy to Eq.~\eqref{eq:C1-1st-lowT}. The collective nature of the interaction, last term in Eq.~\eqref{eq:1st-hamtot-N}, is in the resulting coherence \eqref{eq:C1-1st-lowT-N} reflected by the accumulated value of the respective coupling constants $\gamma_j$. Therefore, it can advantageously reach the same coherence as in \eqref{eq:C1-1st-lowT}, 
using many TLSs with smaller couplings $\gamma_j$, instead of a single strongly coupled TLS, see Fig.~\ref{fig:coh-bath-vs-2nd}(b). This value can be obtained if $\sum_j^{N+1}\gamma_j\gg \omega_1$, supporting the hypothesis that the structure of ground state is in a separable form $\ket{E_0}=(\cos\epsilon_{0,N}\ket{e_1}+\sin\epsilon_{0,N}\ket{g_1})\bigotimes_j\ket{g_j}$. The comparison of numerical results for several values of $N$ is shown in Appendix~\ref{sec:S-enhancement}, Fig.~\ref{sfig:coh-2nd-N}. 

 Analogously as in Sec.~\ref{sub-direct} we can apply the generated coherence to the phase estimation and quantify the estimation error by an inverse of quantum Fisher information~\cite{BrivioPRA2010}, which we label $H_N$ in the case of $N$ ancillary TLSs, Eq.~\eqref{eq:C1-1st-lowT-N}. It amounts to 
\begin{equation}
H_N=\left(C_1^N\right)^2=\frac{\left(\sum_{j=2}^{N+1}\gamma_j\right)^2}{\left[\left(\sum_{j=2}^{N+1}\gamma_{j}\right)^2+\omega_1^2\right]}.
\label{eq-QFI-N}
\end{equation}
As quantum coherence can increase by adding $N$ ancillary TLSs, quantum Fisher information increases correspondingly, demonstrating suppression of the estimation error ($\propto 1/H_N$).   

Similar analysis can be performed in the case of generalized indirect coherence $C_{\rm T}^N$ generation on a target "T" TLS by $N$ source TLS, see~\eqref{eq:2nd-hamtot} and  \eqref{eq:2nd-hamtot-N}, yielding the low temperature approximation of the coherence $C_{\rm T}^N$
\begin{eqnarray}
C_{\rm T}^N\approx \frac{N\gamma^2}{\omega_1(\omega_{\rm T}+\omega_1)},\quad T\ll N\gamma^2\ll\;\omega_{1({\rm T})},
\label{eq:2nd-lowT-N}
\end{eqnarray}
under assumption $\gamma_j\approx\theta_j =\gamma$, $N\geq 1$. Numerical analysis reveals no meaningful possibility of reaching $C_{\rm T}^N\gtrsim C_1^N$, see~Appendix~\ref{sec:S-enhancement}, Fig.~\ref{sfig:coh-2nd-N}.

Finally we can analyze the coherence generated directly by $N$ sources and transferred to the target system "0". In the limit of low temperature assuming $\omega_j= \omega_2$ and $\gamma_j=\gamma,\,j\geq 2$, see~Fig.~\ref{sfig:coh-2nd-BS-N}, we obtain 
\begin{eqnarray}
C_0^N\approx \frac{N\theta\,\gamma}{\sqrt{(N\theta\,\gamma)^2+\omega_0^2[(N\gamma)^2+\omega_1^2]}},
\label{eq:C1-2nd-BS-lowT-N}
\end{eqnarray}
assuming $T\ll \gamma,\theta,\omega_{0(1,2)},\, N\geq 1$. Similarly as in the discussion below Eq.~\eqref{eq:C1-2nd-BS-lowT}, we notice that in the low $T$ limit $C_0^N/C_1^N=x/\sqrt{1+[C_1^N]^2x^2}$, $x=\theta/\omega_0$, allowing for enhancement of low $C_1^N$ coherence values, yielding ratio $C_0^N/C_1^N >1$ for $x\geq (1-[C_1^N]^2)^{-1/2}$. 

\section{Conclusion} We have considered a system of finite number of mutually interacting two level systems (TLSs) brought into a global Gibbs state by the presence of a thermal bath. Based on the symmetry arguments, we have formulated a necessary condition for autonomous generation of non-zero local coherence of the individual TLS. In the case of a pair of TLS, the condition is shown to be sufficient, as well. Such autonomous appearance of local coherence is illustrated on several examples stemming from quantum models describing platform of superconducting circuits~\cite{BlaisPRA2004Transmon1,KochPRA2007Transmon2,Hamedani_Pekola_Entropy2021,LisenfeldPRL2010,
Hamedani_Pekola_Entropy2021,PekolaRevModPhys2021,GuthriePhysRevApp2022}. In principle, such platform should offer as well the possibility to effectively build-up strong coupling effects useful for autonomous coherence generation. It is qualitatively more feasible and extendable approach than engineering of system-bath interaction in the original proposals~\cite{giacomoPRL2018,GUARNIERIPLA2020,purkayastha2020tunable,AndersPRL2021strongGibbs,
slobodeniuk2021extraction,RicardoPRA2021}. 

From quantitative point of view, the amount of coherence generated autonomously at low enough temperatures in our settings can be higher compared to the original design \cite{giacomoPRL2018} based on the system-bath coupling engineering, see Fig.~\ref{fig:coh-bath-vs-2nd}(a), under comparable conditions and parameter settings. Some settings that we examine in our work allow for (generally in the strong coupling regime between TLSs) reaching high values of the generated coherence, in principle approaching the maximum value for a single qubit ($C=1$). If the strong coupling between the subsystems is not available, we show that corresponding effects can be effectively built-up by collective interaction of larger number of ancillary TLSs with the TLS of interest, see Sec.~\ref{sec-N} and Fig.~\ref{fig:coh-bath-vs-2nd}(b). 
Having the autonomously generated TLS coherence available, one can utilize it for adequate tasks, e.g. for phase-estimation process. It makes the first step of Ramsey single-atom interferometry autonomous, being a necessary step towards a new class of autonomous quantum sensors. 
Our work can further trigger future discussion on thermodynamic applications, such as quantum heat engines utilizing autonomous coherence, or on preparation of states with quantum properties for quantum information tasks, in the presence of noisy (thermal) environment, e.g., in the case of computing with noisy intermediate-scale quantum devices~\cite{Preskill2018quantumcomputing,GuzikRevModPhys2022}. 

In future, the study of interplay between the system-bath and intra-system SSC-generating interactions when the former is not considered negligible, might be of fundamental interest. In particular, it would be interesting to see how the conditions identified in this work would be modified by considering the system as given by a mean-force Gibbs state rather than a thermal state. A non trivial competition between these two effects might in principle lead to enhanced SSC.

\section*{Acknowledgments}  The authors acknowledge the helpful discussions on the manuscript with G. Guarnieri. M.K. and R.F. acknowledge support through Project No. 22-27431S of the Czech Science Foundation and the European Union’s 2020 research and innovation
programme (CSA - Coordination and support action,
H2020-WIDESPREAD-2020-5) under grant agreement
No. 951737 (NONGAUSS).

\appendix

\section{Symmetry of local-coherence non-generating Hamiltonians}
\label{sec:S-symmetry}
We consider the so called $\mathbb{Z}_2$ symmetry, inverting $x,y$ spin components
\begin{eqnarray}
\mathbb{Z}_2:\;\op{\sigma}^{x(y)}_{i}\rightarrow -\op{\sigma}^{x(y)}_{i},\quad \op{\sigma}^{z}_{i}\rightarrow\op{\sigma}^{z}_{i}. 
\label{eq:Z2-symmetry}
\end{eqnarray}
In a sense, $\mathbb{Z}_2$ can be regarded as an inversion with respect to $z$-axis (or phase-flip). 
The symmetry in Eq.~\eqref{eq:Z2-symmetry} is generated by 
\begin{eqnarray}
\op{\mathbb{Z}}=\bigotimes_i\op{\sigma}_{i}^z,
\label{eq:Z2-symmetry-gen}
\end{eqnarray}
where the generator $\op{\mathbb{Z}}$ eigenstates define the basis with respect to which we quantify the coherence. 

{\color{black} If the Hamiltonian $\op{H}^{tot}_{S}(\kappa)$, specified by the set of parameters $\kappa$, has $\mathbb{Z}_2$ symmetry meaning $\op{\mathbb{Z}}\op{H}^{tot}_{S}(\kappa)\op{\mathbb{Z}}^\dagger=\op{H}^{tot}_{S}(\kappa)$, then the Gibbs (thermal) state $\op{\tau}$, Eq.~\eqref{eq:gibbs-tot}, defined by such Hamiltonian has the same symmetry. In this case 
\begin{eqnarray}\nonumber
{\rm Tr}[\op{\tau}\op{\sigma}^{x(y)}_{i}]&=&{\rm Tr}[(\op{\mathbb{Z}}\op{\tau}\op{\mathbb{Z}}^\dagger)\op{\sigma}^{x(y)}_{i}]={\rm Tr}[\op{\tau}(\op{\mathbb{Z}}^\dagger\op{\sigma}^{x(y)}_{i}\op{\mathbb{Z}})]\\
&=&-{\rm Tr}[\op{\tau}\op{\sigma}^{x(y)}_{i}]\Rightarrow {\rm Tr}[\op{\tau}\op{\sigma}^{x(y)}_{i}]\equiv 0.
\label{eq:coherence-vanish}
\end{eqnarray}
Equation~\eqref{eq:coherence-vanish} implies {\it zero local coherence}, as $C_j(\kappa)\equiv |\langle\op{\sigma}_j^x+i\op{\sigma}_j^y\rangle_{\op{\tau}}|$, $\langle\bullet\rangle_{\op{\tau}}\equiv {\rm Tr}(\bullet\op{\tau})$. Hence, we can formulate the {\it necessary} condition for thermal coherence generation
\begin{eqnarray}
[\op{\mathbb{Z}},\op{H}^{tot}_{S}(\kappa)]=0\,\Rightarrow\,\forall i,\kappa\quad C_i(\kappa)=0.
\label{eq:coherence-necessary}
\end{eqnarray}
We can not prove the reversed implication in full generality, as zero local coherence $C_i(\kappa)=0$ implies only the equality ${\rm Tr}[(\op{\mathbb{Z}}\op{\tau}\op{\mathbb{Z}}^\dagger)\op{\sigma}^{x(y)}_{i}]={\rm Tr}[\op{\tau}\op{\sigma}^{x(y)}_{i}]$, from which does not follow the symmetry $\op{\mathbb{Z}}\op{H}^{tot}_{S}(\kappa)\op{\mathbb{Z}}^\dagger=\op{H}^{tot}_{S}(\kappa)$ of the Hamiltonian generating the Gibbs state~\eqref{eq:gibbs-tot}. 

Although generally too complex, the proof can be done in the simplest nontrivial case of a pair of TLS. Performing the proof by contradiction, we assume to be simultaneously true
\begin{eqnarray}
\forall i,\kappa\quad C_i(\kappa)=0\quad\,and\quad\op{\mathbb{Z}}\op{H}^{tot}_{S}(\kappa)\op{\mathbb{Z}}\neq\op{H}^{tot}_{S}(\kappa),
\label{eq:S-proof-contradiction}
\end{eqnarray}
and without loss of generality assuming no presence of local terms of the type $\op{\sigma}^{x(y)}_{1(2)}$ in $\op{H}^{tot}_{S}(\kappa)$. The asymmetry of $\op{H}^{tot}_{S}(\kappa)$ then implies the presence of, e.g., the interaction term $\op{\sigma}^x_{1}\op{\sigma}^z_{2}$, together with the terms of the form $\op{\sigma}^j_{1}\op{\sigma}^j_{2}$, with $j=x,y,z$. But, as can be checked by direct inspection, such form of Hamiltonian {\it generates} local coherence in the Gibbs state, i.e. $\exists\, i,\kappa:\,C_i(\kappa)\neq 0$, contradicting the initial assumption. Therefore, for a pair of interacting TLS we have proved the equivalence
\begin{eqnarray}
[\op{\mathbb{Z}},\op{H}^{tot}_{S}(\kappa)]=0\,\Leftrightarrow\,\forall i,\kappa\quad C_i(\kappa)= 0.
\label{eq:coherence-equiv}
\end{eqnarray}
}

The equivalence in Eq.~\eqref{eq:coherence-equiv} can be straightforwardly demonstrated on the example of general XYZ two-spin chain Hamiltonian 
\begin{eqnarray}\nonumber
\op{H}^{tot}_{S}&=&\frac{\omega_1}{2}\op{\sigma}^z_{1}+\frac{\omega_2}{2}\op{\sigma}^z_{2}\\
&+&\left(\frac{\gamma_x}{2}\op{\sigma}^x_{1}\op{\sigma}^x_{2}+\frac{\gamma_y}{2}\op{\sigma}^y_{1}\op{\sigma}^y_{2}+\frac{\gamma_z}{2}\op{\sigma}^z_{1}\op{\sigma}^z_{2}\right),
\label{eq-Htot-XYZ}
\end{eqnarray}
yielding zero local coherence for any values of the interaction $\gamma_\alpha$, $\alpha=x,y,z$, and covering the "RWA" ($\gamma_x=\gamma_y$, $\gamma_z=0$) or "X-X" interactions ($\gamma_y=\gamma_z=0$).

\section{High temperature limits of coherence}
\label{sec:S-high-T}
Let us point out one general feature of any {\it finite dimensional} thermal state $\op{\tau}$, Eq.~\eqref{eq:gibbs-tot}. In the high temperature limit, $T\gg |E_j-E_i|$, with $E_k$ being the $\op{H}_S^{tot}$ eigenvalues, the Gibbs state can be approximated as $\op{\tau}\approx \op{\mathbb{1}}/D$, $D$ being the system dimension. In this limit, coherence of $\op{\tau}$ vanishes in any meaningful sense.

For the first method (directly induced coherence), the value of coherence for low temperatures~\eqref{eq:C1-1st-lowT} naturally contrasts with coherence $C_1$ behavior in the large temperature $T$ limit. Taking expansion for small arguments $\tanh x\approx x$, Eq.~\eqref{eq:C1-1st} can be approximated as
\begin{eqnarray}
C_1\approx \frac{\gamma\omega_2}{4\,T^{2}},\quad T\gg \gamma,\omega_{1(2)},
\label{eq:C1-1st-highT}
\end{eqnarray}
showing that the coherence vanishes in the classical (high $T$) limit, confirming the general conclusion mentioned above, thus complying with the intuition that quantum properties should vanish in classical (high $T$) limit, as expected~\cite{giacomoPRL2018, AndersPRL2021strongGibbs,TrushechkinQuSci2022}. Complementary, this asymptotic behavior is independent of the frequency $\omega_1$ of the TLS of interest. 

In the next method (indirectly induced coherence) considered in the main text, the asymptotic expansion for large temperatures $T$ yields
\begin{eqnarray}
C_2\approx \frac{\gamma\theta\omega_2}{24\,T^{3}},\quad T\gg \gamma,\;\theta,\;\omega_{1(2)}.
\label{eq:C1-2nd-highT}
\end{eqnarray}
Interestingly, one can note several differences compared to Eq.~\eqref{eq:C1-1st-highT}. The most notable one is the faster asymptotic approach to zero value of coherence $C_2$ with $T$, due to the inverse cubic $T$ dependence as opposed to inverse quadratic one in Eq.~\eqref{eq:C1-1st-highT}. Another interesting aspect of the high temperature limit of $C_2$ \eqref{eq:C1-2nd-highT} is its independence on the parameters characterizing the first TLS. Moreover, the $\gamma\,\theta$ dependence is of the {\it second} (higher) order compared to the direct case~\eqref{eq:C1-1st-highT}, thus resulting in much weaker, although still observable effect of coherence generation on the second TLS.

In the last scenario (transferred directly induced coherence), considered in this work, the high temperature limit obtained for the target TLS "0" $C_0$ yields
\begin{eqnarray}
C_0\approx \frac{\gamma\theta\omega_2}{8\,T^{3}},\; T\gg \gamma,\,\theta,\,\omega_{0(1,2)},
\label{eq:C1-2nd-BS-highT}
\end{eqnarray}
where the coherence remarkably depends only on the frequency of the second TLS, cf. Eq.~\eqref{eq:C1-2nd-highT}, and keeps inverse cubic dependency on temperature $T$, as the indirect method. It shows that the high-temperature regime does not necessarily distinguish the properties of coherence generation, revealed in the low temperature limit.

\section{Coherence generation by multiple ancillary TLS}
\label{sec:S-enhancement}
\begin{figure*}[ht]
\begin{tikzpicture} 
  \node (img1)  {\includegraphics[width=.9\columnwidth]{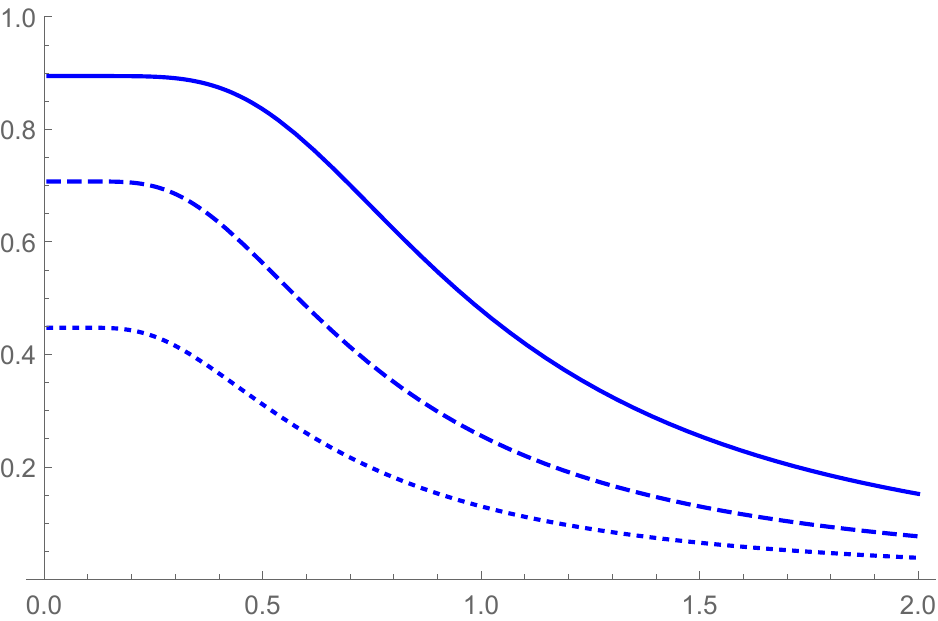}};
  \node[above=of img1, node distance=0cm, yshift=-2.3cm,xshift=1.7cm] {{\color{black}{\bf (a)}}};
  \node[above=of img1, node distance=0cm, yshift=-1.7cm,xshift=.5cm] {Direct\;coherence\;$C_1^N$\;generation\;from $N$\;TLSs};
  \node[above=of img1, node distance=0cm, yshift=-6.9cm,xshift=0cm] {$T/\omega_1$};
   \node[above=of img1, node distance=0cm, yshift=-4.5cm,xshift=-2.8cm]{{\color{blue}$N=1$}};\node[above=of img1, node distance=0cm, yshift=-3.3cm,xshift=-2.8cm]{{\color{blue}$N=2$}};\node[above=of img1, node distance=0cm, yshift=-2.5cm,xshift=-.5cm]{{\color{blue}$N=4$}};\node[above=of img1, node distance=0cm, yshift=-3.8cm,xshift=2.8cm]{{\color{blue}$\gamma_j=0.5$}};
   \node[above=of img1, node distance=0cm, yshift=-3cm,xshift=2.8cm] {{\color{blue}$\omega_j\equiv\omega_2=1.3$}};
    \node[left=of img1, node distance=0cm, rotate=90, anchor=center, yshift=-.7cm,xshift=-0.cm] {{\color{black}$C_1^N$}};
\end{tikzpicture}
\hfill
\begin{tikzpicture} 
  \node (img1)  {\includegraphics[width=.9\columnwidth]{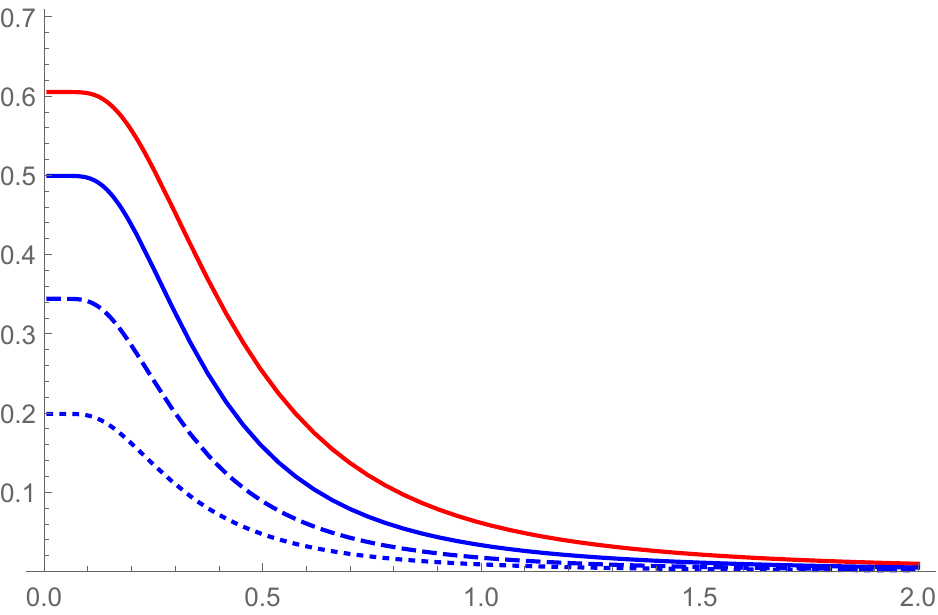}};
  \node[above=of img1, node distance=0cm, yshift=-1.7cm,xshift=.5cm] {Indirect\;coherence\;$C_{\rm T}^N$\;generation\;from $N$\;TLSs};
  \node[above=of img1, node distance=0cm, yshift=-2.3cm,xshift=1.7cm] {{\color{black}{\bf (b)}}};
  \node[above=of img1, node distance=0cm, yshift=-6.8cm,xshift=0cm] {$T/\omega_{\rm T}$};
   \node[above=of img1, node distance=0cm, yshift=-5.7cm,xshift=-3cm]{{\color{blue}$N=1$}};
   \node[above=of img1, node distance=0cm, yshift=-3.3cm,xshift=-1.5cm]{{\color{blue}$N=4$}};
   \node[above=of img1, node distance=0cm, yshift=-2.1cm,xshift=-2.cm]{{\color{red}$N=8$}};
   \node[above=of img1, node distance=0cm, yshift=-3.5cm,xshift=2.8cm]{{\color{blue}$\theta_j=\gamma_j\equiv\gamma=0.5$}};
   \node[above=of img1, node distance=0cm, yshift=-4.1cm,xshift=2.8cm] {{\color{blue}$\omega_j\equiv\omega_1=0.5$}};
    \node[left=of img1, node distance=0cm, rotate=90, anchor=center, yshift=-.7cm,xshift=-0.cm] {{\color{black}$C_{\rm T}^N$}};
\end{tikzpicture}
\caption{The dependence of coherence $C_{1({\rm T})}^N$ generated by $N$ TLSs on $T$ for various values of parameters. (a) The direct coherence generation determined by Hamiltonian~\eqref{Seq:1st-hamtot-N}. In all cases the choice $\omega_1=1$ was made. Note that the width of the low $T$ plateau increases with increasing $N$. (b)  Indirect coherence generation, Hamiltonian~\eqref{eq:2nd-hamtot-N}. The values of the parameters were set to $\omega_{\rm T}=1$, $\theta_j=\gamma_j=0.5$, and $\omega_j=0.5$. The red curve represents the results for $N=8$ ancillary TLSs. It approximately represents an edge of our numerical possibilities. As well, it is close to the $C_{{\rm T}}^N$ saturation curve for given values of the parameters, i.e. further increase of $N$ will change $C_{{\rm T}}^N$ values only negligibly.    }
\label{sfig:coh-2nd-N}
\end{figure*}
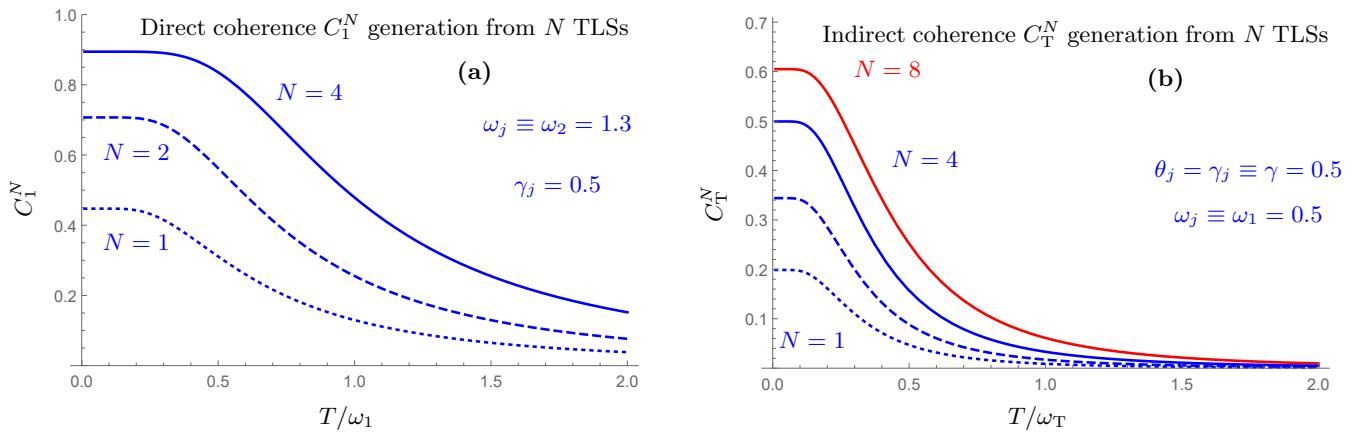

In this section, we discuss the effect of coupling $N$ {\it different} source-TLSs to our system of interest via the interaction Hamiltonians listed in previous sections. We will focus our attention again on the limiting cases of high and low bath temperatures $T$. Let us start the discussion with the direct case. The generalized total Hamiltonian determining the thermal state is
\begin{eqnarray}
\op{H}^{tot}_{S}=\frac{\omega_1}{2}\;\op{\sigma}_1^z+\sum_{j=2}^{N+1}\frac{\omega_j}{2}\;\op{\sigma}_j^z+\sum_{j=2}^{N+1}\frac{\gamma_j}{2}\;\op{\sigma}_1^x\op{\sigma}_j^z,
\label{Seq:1st-hamtot-N}
\end{eqnarray}
with $\gamma_j >0,\;N\geq 1$. The full analysis can be done numerically in a straightforward way. We proceed further with less generality but with possibility of obtaining more tractable analytical results by assuming that  all source TLS are approximately identical, $\omega_j\approx\omega_2$. Such approximation allows us to see the generally {\it positive} trend of increasing the number $N$ of source TLSs. In the high $T$ limit, the expectation of vanishing coherence holds, as the target system coherence $C_1^N$ reads
\begin{eqnarray}
C_1^N\approx \frac{\sum_j^{N+1}\gamma_j\omega_2}{4T^{2}},\quad T\gg \gamma_j,\omega_{1(2)},
\label{eq:C1-1st-highT-N}
\end{eqnarray}
revealing that the general asymptotic dependence of coherence on $T^{-2}$ is still present, see Eq.~\eqref{eq:C1-1st-highT}. We can also conclude, that coupling of $N$ TLS simultaneously amounts to an effective increase of the interaction constant $\gamma\rightarrow \sum_j^{N+1}\gamma_j$, by comparison of Eqs.~\eqref{eq:C1-1st-highT}~and~\eqref{eq:C1-1st-highT-N}. Similar observation can be also made in the opposite low $T$ limit for the coherence $C_1^N$, again assuming the source TLSs being approximately identical, yielding
\begin{eqnarray}
C_1^N\approx \frac{\sum_j^{N+1}\gamma_j}{\sqrt{(\sum_j^{N+1}\gamma_j)^2+\omega_1^2}},
\label{Seq:C1-1st-lowT-N}
\end{eqnarray}
where $T\ll \gamma_j,\omega_{1(2)},\, N\geq 1$, showing again independence on the source frequency $\omega_2$, see~\eqref{eq:C1-1st-lowT}. Another interesting property stemming from Eq.~\eqref{Seq:C1-1st-lowT-N} is, that it can in principle still reach the absolute optimum $C_1^N\approx 1$. This value can be obtained if $\sum_j^{N+1}\gamma_j\gg \omega_1$, supporting the hypothesis about the structure of ground state to be in a separable form $\ket{E_0}=(\cos\alpha_{0,N}\ket{e_1}+\sin\alpha_{0,N}\ket{g_1})\bigotimes_j\ket{g_j}$. The comparison of numerical results for several values of $N$ is shown in Fig.~\ref{sfig:coh-2nd-N}. 

Similar analysis can be performed in the case of indirect coherence $C_{\rm T}^N$ generation by $N$ source TLS, where "T" labels the {\it target} system of interest. The total Hamiltonian reads
\begin{eqnarray}\nonumber
\op{H}^{tot}_{S}&=&\frac{\omega_T}{2}\;\op{\sigma}_{\rm T}^z+\sum_{j=1}^{N}\frac{\omega_j}{2}\;\op{\sigma}_j^z\\
&+&\sum_{j=1}^{N}\left(\frac{\gamma_j}{2}\;\op{\sigma}_{\rm T}^z\op{\sigma}_j^x+\frac{\theta_j}{2}\;\op{\sigma}_{\rm T}^x\op{\sigma}_j^x\right),\, 0<\gamma_j,\theta_j.
\label{eq:2nd-hamtot-N}
\end{eqnarray}
For the sake of simplicity, but without loss of generality, we directly assume in this case $\gamma_j\approx\theta_j\approx\gamma$ and $\omega_j\approx\omega_1$. With these assumptions, the high $T$ asymptotic behavior can be approximated as 
\begin{eqnarray}
C_{\rm T}^N\approx \frac{N\gamma^2\omega_T}{24\,T^{3}},\quad T\gg \gamma,\;\omega_{1(2)}.
\label{eq:C1-2nd-highT-N}
\end{eqnarray}
Under the same assumptions, we can obtain the low $T$ approximation of the coherence $C_{\rm T}^N$
\begin{eqnarray}
C_{\rm T}^N\approx \frac{N\gamma^2}{\omega_1(\omega_{\rm T}+\omega_1)},\quad T\ll N\gamma\ll\;\omega_{1({\rm T})}.
\label{Seq:2nd-lowT-N}
\end{eqnarray}
This approximate result valid for low temperatures and low couplings was estimated based on numerical evidence and checked up to $N=3$, cf. Fig.~\ref{sfig:coh-2nd-N}. It is important to stress, that it represent only the leading term of the generated coherence $C_{\rm T}^N$. 

Finally we can analyze the coherence generated directly by $N$ sources and transferred to the target system, in the spirit of transferred directly induced method. Such situation can be described by the extended (with respect to~\eqref{eq:2nd-BS-hamtot}) Hamiltonian 
\begin{eqnarray}\nonumber
\op{H}^{tot}_{S}&=&\frac{\omega_0}{2}\;\op{\sigma}_0^z+\frac{\theta}{2}\;\op{\sigma}_0^x\op{\sigma}_1^x+\frac{\omega_1}{2}\;\op{\sigma}_1^z\\
&+&\sum_{j=2}^{N+1}\frac{\gamma_j}{2}\;\op{\sigma}_1^x\op{\sigma}_j^z+\sum_{j=2}^{N+1}\frac{\omega_j}{2}\;\op{\sigma}_j^z,
\label{eq:2nd-BS-hamtot-N}
\end{eqnarray}
where $\gamma_j,\theta>0,\, N\geq 1$.
Although the analysis of the full model, with general values of $\gamma_j$, $\omega_j$ might be interesting, we focus our attention to a simplified case $\omega_j=\omega_2$ and $\gamma_j=\gamma,\,j\geq 2$. Such simplification allows for obtaining a semi-analytic result valid in the high temperature limit 
\begin{eqnarray}
C_0^N\approx A/ T^{3}\quad T\gg \gamma_j,\theta,\omega_{0(1,j)},
\label{eq:C1-2nd-BS-highT-N}
\end{eqnarray}
where we leave the coefficient $A$ unspecified due to its complicated structure.

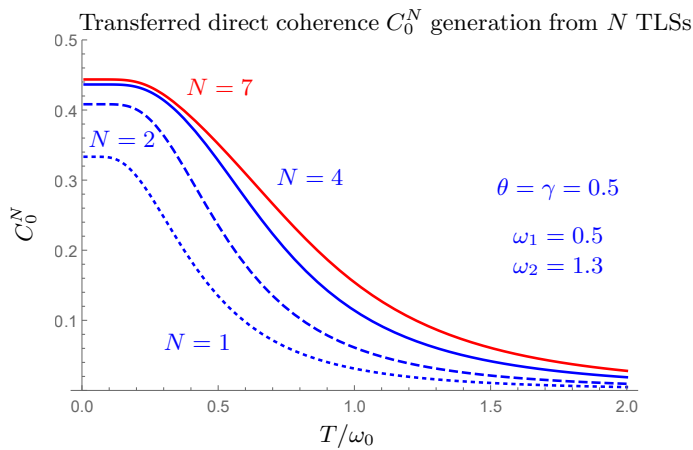
\begin{figure*}[ht]
\begin{tikzpicture} 
  \node (img1)  {\includegraphics[width=.9\columnwidth]{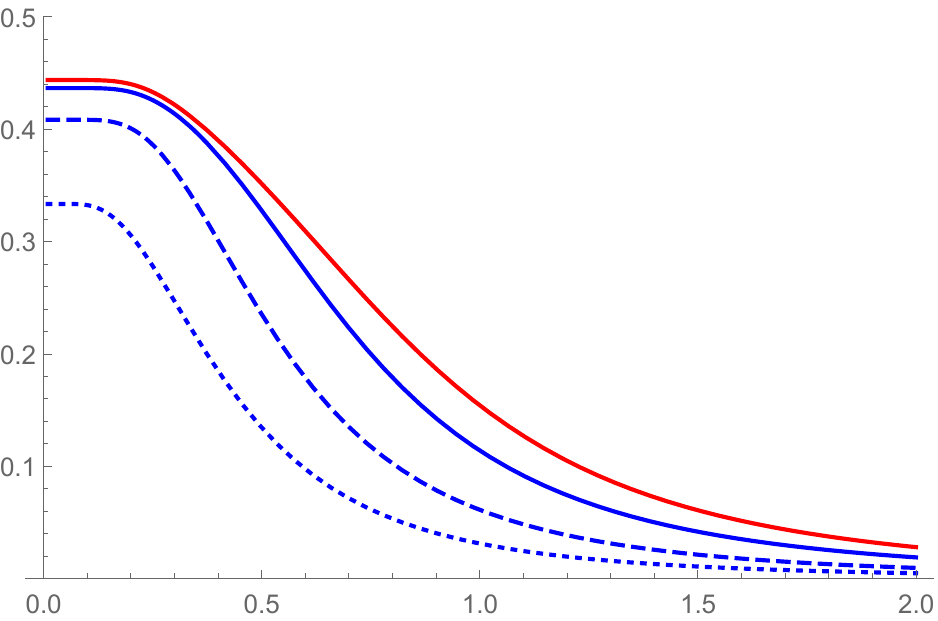}};
  \node[above=of img1, node distance=0cm, yshift=-1.3cm,xshift=.5cm] {Transferred\;direct\;coherence\;$C_0^N$\;generation\;from $N$\;TLSs};
  \node[above=of img1, node distance=0cm, yshift=-6.8cm,xshift=0cm] {$T/\omega_0$};
   \node[above=of img1, node distance=0cm, yshift=-5.5cm,xshift=-2cm]{{\color{blue}$N=1$}};
   \node[above=of img1, node distance=0cm, yshift=-2.8cm,xshift=-3cm]{{\color{blue}$N=2$}};
   \node[above=of img1, node distance=0cm, yshift=-3.3cm,xshift=-.5cm]{{\color{blue}$N=4$}};
   \node[above=of img1, node distance=0cm, yshift=-2.1cm,xshift=-1.7cm]{{\color{red}$N=7$}};
   \node[above=of img1, node distance=0cm, yshift=-3.5cm,xshift=2.8cm]{{\color{blue}$\theta=\gamma=0.5$}};
   \node[above=of img1, node distance=0cm, yshift=-4.5cm,xshift=2.8cm]{{\color{blue}$\omega_2=1.3$}};
   \node[above=of img1, node distance=0cm, yshift=-4.1cm,xshift=2.8cm] {{\color{blue}$\omega_1=0.5$}};
    \node[left=of img1, node distance=0cm, rotate=90, anchor=center, yshift=-.7cm,xshift=-0.cm] {{\color{black}$C_0^N$}};
\end{tikzpicture}
\caption{The temperature dependence of coherence $C_0^N$ generated by $N$ TLSs and transferred to the target system for various values of $N$ according to the Hamiltonian~\eqref{eq:2nd-BS-hamtot-N}. In all cases the choice $\omega_0=1$ was made. The values of the parameters were set to $\omega_3=1.3$, $\theta=\gamma_3=0.5$, and $\omega_2=0.5$. The red curve represents the results for $N=7$ ancillary TLS. It approximately represents an edge close to saturation of $C_0^N$ curve for given values of the parameters, i.e. further increase of $N$ will change $C_0^N$ values only negligibly. This intuitively points out the limited capacity of the single  $\theta$-coupling term in~\eqref{eq:2nd-BS-hamtot-N} for coherence transfer.  }
\label{sfig:coh-2nd-BS-N}
\end{figure*}

In the opposite, and more attractive, limit of low temperature $T\ll \gamma,\theta,\omega_{1(2)}$ and $N\geq 1$, we obtain the value
\begin{eqnarray}
C_0^N\approx \frac{\theta\,N\gamma}{\sqrt{(\theta\,N\gamma)^2+\omega_0^2[(N\gamma)^2+\omega_1^2]}}.
\label{Seq:C1-2nd-BS-lowT-N}
\end{eqnarray}
This result for the coherence of the target system, transferred from the mediator system, where it is generated by $N$ TLSs can be viewed as an interesting composition of our previous results, cf. Eqs.~\eqref{eq:C1-2nd-BS-lowT}~and~\eqref{Seq:C1-1st-lowT-N}, revealing generalized "transfer" rule mentioned below Eq.~\eqref{eq:C1-2nd-BS-lowT}, now with $N\gamma\rightarrow\theta N\gamma$ and $\omega_1\rightarrow\omega_0\sqrt{(N\gamma)^2+\omega_1^2}$. We have numerically presented the result \eqref{Seq:C1-2nd-BS-lowT-N} up to $N=7$ in Fig.~\ref{sfig:coh-2nd-BS-N}. We can intuitively recognize the limited  capacity of the coherence-transferring channel $\op{\sigma}_0^x\op{\sigma}_1^x$, leading to a saturation of the coherence $C_0^N$ already for relatively small value of $N\approx 4$. 
  
\bibliography{refs}

\begin{thebibliography}{54}%
\makeatletter
\providecommand \@ifxundefined [1]{%
 \@ifx{#1\undefined}
}%
\providecommand \@ifnum [1]{%
 \ifnum #1\expandafter \@firstoftwo
 \else \expandafter \@secondoftwo
 \fi
}%
\providecommand \@ifx [1]{%
 \ifx #1\expandafter \@firstoftwo
 \else \expandafter \@secondoftwo
 \fi
}%
\providecommand \natexlab [1]{#1}%
\providecommand \enquote  [1]{``#1''}%
\providecommand \bibnamefont  [1]{#1}%
\providecommand \bibfnamefont [1]{#1}%
\providecommand \citenamefont [1]{#1}%
\providecommand \href@noop [0]{\@secondoftwo}%
\providecommand \href [0]{\begingroup \@sanitize@url \@href}%
\providecommand \@href[1]{\@@startlink{#1}\@@href}%
\providecommand \@@href[1]{\endgroup#1\@@endlink}%
\providecommand \@sanitize@url [0]{\catcode `\\12\catcode `\$12\catcode
  `\&12\catcode `\#12\catcode `\^12\catcode `\_12\catcode `\%12\relax}%
\providecommand \@@startlink[1]{}%
\providecommand \@@endlink[0]{}%
\providecommand \url  [0]{\begingroup\@sanitize@url \@url }%
\providecommand \@url [1]{\endgroup\@href {#1}{\urlprefix }}%
\providecommand \urlprefix  [0]{URL }%
\providecommand \Eprint [0]{\href }%
\providecommand \doibase [0]{https://doi.org/}%
\providecommand \selectlanguage [0]{\@gobble}%
\providecommand \bibinfo  [0]{\@secondoftwo}%
\providecommand \bibfield  [0]{\@secondoftwo}%
\providecommand \translation [1]{[#1]}%
\providecommand \BibitemOpen [0]{}%
\providecommand \bibitemStop [0]{}%
\providecommand \bibitemNoStop [0]{.\EOS\space}%
\providecommand \EOS [0]{\spacefactor3000\relax}%
\providecommand \BibitemShut  [1]{\csname bibitem#1\endcsname}%
\let\auto@bib@innerbib\@empty
\bibitem [{\citenamefont {Streltsov}\ \emph {et~al.}(2017)\citenamefont
  {Streltsov}, \citenamefont {Adesso},\ and\ \citenamefont
  {Plenio}}]{StreltsovRevModPhys2017}%
  \BibitemOpen
  \bibfield  {author} {\bibinfo {author} {\bibfnamefont {A.}~\bibnamefont
  {Streltsov}}, \bibinfo {author} {\bibfnamefont {G.}~\bibnamefont {Adesso}},\
  and\ \bibinfo {author} {\bibfnamefont {M.~B.}\ \bibnamefont {Plenio}},\
  }\bibfield  {title} {\bibinfo {title} {Colloquium: Quantum coherence as a
  resource},\ }\href {https://doi.org/10.1103/RevModPhys.89.041003} {\bibfield
  {journal} {\bibinfo  {journal} {Rev. Mod. Phys.}\ }\textbf {\bibinfo {volume}
  {89}},\ \bibinfo {pages} {041003} (\bibinfo {year} {2017})}\BibitemShut
  {NoStop}%
\bibitem [{\citenamefont {Chitambar}\ and\ \citenamefont
  {Gour}(2019)}]{ChitambarRevModPhys2019}%
  \BibitemOpen
  \bibfield  {author} {\bibinfo {author} {\bibfnamefont {E.}~\bibnamefont
  {Chitambar}}\ and\ \bibinfo {author} {\bibfnamefont {G.}~\bibnamefont
  {Gour}},\ }\bibfield  {title} {\bibinfo {title} {Quantum resource theories},\
  }\href {https://doi.org/10.1103/RevModPhys.91.025001} {\bibfield  {journal}
  {\bibinfo  {journal} {Rev. Mod. Phys.}\ }\textbf {\bibinfo {volume} {91}},\
  \bibinfo {pages} {025001} (\bibinfo {year} {2019})}\BibitemShut {NoStop}%
\bibitem [{\citenamefont {Regula}\ \emph {et~al.}(2018)\citenamefont {Regula},
  \citenamefont {Fang}, \citenamefont {Wang},\ and\ \citenamefont
  {Adesso}}]{RegulaPRL2018}%
  \BibitemOpen
  \bibfield  {author} {\bibinfo {author} {\bibfnamefont {B.}~\bibnamefont
  {Regula}}, \bibinfo {author} {\bibfnamefont {K.}~\bibnamefont {Fang}},
  \bibinfo {author} {\bibfnamefont {X.}~\bibnamefont {Wang}},\ and\ \bibinfo
  {author} {\bibfnamefont {G.}~\bibnamefont {Adesso}},\ }\bibfield  {title}
  {\bibinfo {title} {One-shot coherence distillation},\ }\href
  {https://doi.org/10.1103/PhysRevLett.121.010401} {\bibfield  {journal}
  {\bibinfo  {journal} {Phys. Rev. Lett.}\ }\textbf {\bibinfo {volume} {121}},\
  \bibinfo {pages} {010401} (\bibinfo {year} {2018})}\BibitemShut {NoStop}%
\bibitem [{\citenamefont {Fang}\ \emph {et~al.}(2018)\citenamefont {Fang},
  \citenamefont {Wang}, \citenamefont {Lami}, \citenamefont {Regula},\ and\
  \citenamefont {Adesso}}]{adessoPRL2018}%
  \BibitemOpen
  \bibfield  {author} {\bibinfo {author} {\bibfnamefont {K.}~\bibnamefont
  {Fang}}, \bibinfo {author} {\bibfnamefont {X.}~\bibnamefont {Wang}}, \bibinfo
  {author} {\bibfnamefont {L.}~\bibnamefont {Lami}}, \bibinfo {author}
  {\bibfnamefont {B.}~\bibnamefont {Regula}},\ and\ \bibinfo {author}
  {\bibfnamefont {G.}~\bibnamefont {Adesso}},\ }\bibfield  {title} {\bibinfo
  {title} {Probabilistic distillation of quantum coherence},\ }\href
  {https://doi.org/10.1103/PhysRevLett.121.070404} {\bibfield  {journal}
  {\bibinfo  {journal} {Phys. Rev. Lett.}\ }\textbf {\bibinfo {volume} {121}},\
  \bibinfo {pages} {070404} (\bibinfo {year} {2018})}\BibitemShut {NoStop}%
\bibitem [{\citenamefont {Wu}\ \emph {et~al.}(2018)\citenamefont {Wu},
  \citenamefont {Hou}, \citenamefont {Zhao}, \citenamefont {Xiang},
  \citenamefont {Li}, \citenamefont {Guo}, \citenamefont {Ma}, \citenamefont
  {He}, \citenamefont {Thompson},\ and\ \citenamefont {Gu}}]{WuPRL2018}%
  \BibitemOpen
  \bibfield  {author} {\bibinfo {author} {\bibfnamefont {K.-D.}\ \bibnamefont
  {Wu}}, \bibinfo {author} {\bibfnamefont {Z.}~\bibnamefont {Hou}}, \bibinfo
  {author} {\bibfnamefont {Y.-Y.}\ \bibnamefont {Zhao}}, \bibinfo {author}
  {\bibfnamefont {G.-Y.}\ \bibnamefont {Xiang}}, \bibinfo {author}
  {\bibfnamefont {C.-F.}\ \bibnamefont {Li}}, \bibinfo {author} {\bibfnamefont
  {G.-C.}\ \bibnamefont {Guo}}, \bibinfo {author} {\bibfnamefont
  {J.}~\bibnamefont {Ma}}, \bibinfo {author} {\bibfnamefont {Q.-Y.}\
  \bibnamefont {He}}, \bibinfo {author} {\bibfnamefont {J.}~\bibnamefont
  {Thompson}},\ and\ \bibinfo {author} {\bibfnamefont {M.}~\bibnamefont {Gu}},\
  }\bibfield  {title} {\bibinfo {title} {Experimental cyclic interconversion
  between coherence and quantum correlations},\ }\href
  {https://doi.org/10.1103/PhysRevLett.121.050401} {\bibfield  {journal}
  {\bibinfo  {journal} {Phys. Rev. Lett.}\ }\textbf {\bibinfo {volume} {121}},\
  \bibinfo {pages} {050401} (\bibinfo {year} {2018})}\BibitemShut {NoStop}%
\bibitem [{\citenamefont {Hofheinz}\ \emph {et~al.}(2009)\citenamefont
  {Hofheinz} \emph {et~al.}}]{HofheinzNat2009}%
  \BibitemOpen
  \bibfield  {author} {\bibinfo {author} {\bibfnamefont {M.}~\bibnamefont
  {Hofheinz}} \emph {et~al.},\ }\bibfield  {title} {\bibinfo {title}
  {Synthesizing arbitrary quantum states in a superconducting resonator},\
  }\href {https://doi.org/10.1038/nature08005} {\bibfield  {journal} {\bibinfo
  {journal} {Nature}\ }\textbf {\bibinfo {volume} {459}},\ \bibinfo {pages}
  {546} (\bibinfo {year} {2009})}\BibitemShut {NoStop}%
\bibitem [{\citenamefont {Gumberidze}\ \emph {et~al.}(2019)\citenamefont
  {Gumberidze}, \citenamefont {Kolář},\ and\ \citenamefont
  {Filip}}]{Gumberidze2019}%
  \BibitemOpen
  \bibfield  {author} {\bibinfo {author} {\bibfnamefont {M.}~\bibnamefont
  {Gumberidze}}, \bibinfo {author} {\bibfnamefont {M.}~\bibnamefont
  {Kolář}},\ and\ \bibinfo {author} {\bibfnamefont {R.}~\bibnamefont
  {Filip}},\ }\bibfield  {title} {\bibinfo {title} {Measurement induced
  synthesis of coherent quantum batteries},\ }\bibfield  {journal} {\bibinfo
  {journal} {Scientific Reports}\ }\textbf {\bibinfo {volume} {9}},\ \href
  {https://doi.org/10.1038/s41598-019-56158-8} {10.1038/s41598-019-56158-8}
  (\bibinfo {year} {2019})\BibitemShut {NoStop}%
\bibitem [{\citenamefont {St{\'{a}}rek}\ \emph {et~al.}(2021)\citenamefont
  {St{\'{a}}rek}, \citenamefont {Mi{\v{c}}uda}, \citenamefont
  {Kol{\'{a}}{\v{r}}}, \citenamefont {Filip},\ and\ \citenamefont
  {Fiur{\'{a}}{\v{s}}ek}}]{Starek2021}%
  \BibitemOpen
  \bibfield  {author} {\bibinfo {author} {\bibfnamefont {R.}~\bibnamefont
  {St{\'{a}}rek}}, \bibinfo {author} {\bibfnamefont {M.}~\bibnamefont
  {Mi{\v{c}}uda}}, \bibinfo {author} {\bibfnamefont {M.}~\bibnamefont
  {Kol{\'{a}}{\v{r}}}}, \bibinfo {author} {\bibfnamefont {R.}~\bibnamefont
  {Filip}},\ and\ \bibinfo {author} {\bibfnamefont {J.}~\bibnamefont
  {Fiur{\'{a}}{\v{s}}ek}},\ }\bibfield  {title} {\bibinfo {title} {Experimental
  demonstration of optimal probabilistic enhancement of quantum coherence},\
  }\href {https://doi.org/10.1088/2058-9565/ac10ef} {\bibfield  {journal}
  {\bibinfo  {journal} {Quantum Science and Technology}\ }\textbf {\bibinfo
  {volume} {6}},\ \bibinfo {pages} {045010} (\bibinfo {year}
  {2021})}\BibitemShut {NoStop}%
\bibitem [{\citenamefont {Huelga}\ and\ \citenamefont
  {Plenio}(2013)}]{huelga2013vibrations}%
  \BibitemOpen
  \bibfield  {author} {\bibinfo {author} {\bibfnamefont {S.~F.}\ \bibnamefont
  {Huelga}}\ and\ \bibinfo {author} {\bibfnamefont {M.~B.}\ \bibnamefont
  {Plenio}},\ }\bibfield  {title} {\bibinfo {title} {Vibrations, quanta and
  biology},\ }\href@noop {} {\bibfield  {journal} {\bibinfo  {journal}
  {Contemporary Physics}\ }\textbf {\bibinfo {volume} {54}},\ \bibinfo {pages}
  {181} (\bibinfo {year} {2013})}\BibitemShut {NoStop}%
\bibitem [{\citenamefont {Cao}\ \emph {et~al.}(2020)\citenamefont {Cao},
  \citenamefont {Cogdell}, \citenamefont {Coker}, \citenamefont {Duan},
  \citenamefont {Hauer}, \citenamefont {Kleinekathöfer}, \citenamefont
  {Jansen}, \citenamefont {Mančal}, \citenamefont {Miller}, \citenamefont
  {Ogilvie}, \citenamefont {Prokhorenko}, \citenamefont {Renger}, \citenamefont
  {Tan}, \citenamefont {Tempelaar}, \citenamefont {Thorwart}, \citenamefont
  {Thyrhaug}, \citenamefont {Westenhoff},\ and\ \citenamefont
  {Zigmantas}}]{CaoScience2020}%
  \BibitemOpen
  \bibfield  {author} {\bibinfo {author} {\bibfnamefont {J.}~\bibnamefont
  {Cao}}, \bibinfo {author} {\bibfnamefont {R.~J.}\ \bibnamefont {Cogdell}},
  \bibinfo {author} {\bibfnamefont {D.~F.}\ \bibnamefont {Coker}}, \bibinfo
  {author} {\bibfnamefont {H.-G.}\ \bibnamefont {Duan}}, \bibinfo {author}
  {\bibfnamefont {J.}~\bibnamefont {Hauer}}, \bibinfo {author} {\bibfnamefont
  {U.}~\bibnamefont {Kleinekathöfer}}, \bibinfo {author} {\bibfnamefont
  {T.~L.~C.}\ \bibnamefont {Jansen}}, \bibinfo {author} {\bibfnamefont
  {T.}~\bibnamefont {Mančal}}, \bibinfo {author} {\bibfnamefont {R.~J.~D.}\
  \bibnamefont {Miller}}, \bibinfo {author} {\bibfnamefont {J.~P.}\
  \bibnamefont {Ogilvie}}, \bibinfo {author} {\bibfnamefont {V.~I.}\
  \bibnamefont {Prokhorenko}}, \bibinfo {author} {\bibfnamefont
  {T.}~\bibnamefont {Renger}}, \bibinfo {author} {\bibfnamefont {H.-S.}\
  \bibnamefont {Tan}}, \bibinfo {author} {\bibfnamefont {R.}~\bibnamefont
  {Tempelaar}}, \bibinfo {author} {\bibfnamefont {M.}~\bibnamefont {Thorwart}},
  \bibinfo {author} {\bibfnamefont {E.}~\bibnamefont {Thyrhaug}}, \bibinfo
  {author} {\bibfnamefont {S.}~\bibnamefont {Westenhoff}},\ and\ \bibinfo
  {author} {\bibfnamefont {D.}~\bibnamefont {Zigmantas}},\ }\bibfield  {title}
  {\bibinfo {title} {Quantum biology revisited},\ }\href
  {https://doi.org/10.1126/sciadv.aaz4888} {\bibfield  {journal} {\bibinfo
  {journal} {Science Advances}\ }\textbf {\bibinfo {volume} {6}},\ \bibinfo
  {pages} {eaaz4888} (\bibinfo {year} {2020})}\BibitemShut {NoStop}%
\bibitem [{\citenamefont {Arute}\ \emph {et~al.}(2019)\citenamefont {Arute},
  \citenamefont {Arya}, \citenamefont {Babbush}, \citenamefont {Bacon},
  \citenamefont {Bardin}, \citenamefont {Barends}, \citenamefont {Biswas},
  \citenamefont {Boixo}, \citenamefont {Brandao}, \citenamefont {Buell} \emph
  {et~al.}}]{arute2019quantum}%
  \BibitemOpen
  \bibfield  {author} {\bibinfo {author} {\bibfnamefont {F.}~\bibnamefont
  {Arute}}, \bibinfo {author} {\bibfnamefont {K.}~\bibnamefont {Arya}},
  \bibinfo {author} {\bibfnamefont {R.}~\bibnamefont {Babbush}}, \bibinfo
  {author} {\bibfnamefont {D.}~\bibnamefont {Bacon}}, \bibinfo {author}
  {\bibfnamefont {J.~C.}\ \bibnamefont {Bardin}}, \bibinfo {author}
  {\bibfnamefont {R.}~\bibnamefont {Barends}}, \bibinfo {author} {\bibfnamefont
  {R.}~\bibnamefont {Biswas}}, \bibinfo {author} {\bibfnamefont
  {S.}~\bibnamefont {Boixo}}, \bibinfo {author} {\bibfnamefont {F.~G.}\
  \bibnamefont {Brandao}}, \bibinfo {author} {\bibfnamefont {D.~A.}\
  \bibnamefont {Buell}}, \emph {et~al.},\ }\bibfield  {title} {\bibinfo {title}
  {Quantum supremacy using a programmable superconducting processor},\ }\href
  {https://doi.org/10.1038/s41586-019-1666-5} {\bibfield  {journal} {\bibinfo
  {journal} {Nature}\ }\textbf {\bibinfo {volume} {574}},\ \bibinfo {pages}
  {505} (\bibinfo {year} {2019})}\BibitemShut {NoStop}%
\bibitem [{\citenamefont {Hangleiter}\ and\ \citenamefont
  {Eisert}(2022)}]{SupremacyReview}%
  \BibitemOpen
  \bibfield  {author} {\bibinfo {author} {\bibfnamefont {D.}~\bibnamefont
  {Hangleiter}}\ and\ \bibinfo {author} {\bibfnamefont {J.}~\bibnamefont
  {Eisert}},\ }\bibfield  {title} {\bibinfo {title} {Computational advantage of
  quantum random sampling},\ }\href@noop {} {\bibfield  {journal} {\bibinfo
  {journal} {arXiv:2206.04079}\ } (\bibinfo {year} {2022})}\BibitemShut
  {NoStop}%
\bibitem [{\citenamefont {Boixo}\ \emph {et~al.}(2018)\citenamefont {Boixo},
  \citenamefont {Isakov}, \citenamefont {Smelyanskiy}, \citenamefont {Babbush},
  \citenamefont {Ding}, \citenamefont {Jiang}, \citenamefont {Bremner},
  \citenamefont {Martinis},\ and\ \citenamefont
  {Neven}}]{boixo2018characterizing}%
  \BibitemOpen
  \bibfield  {author} {\bibinfo {author} {\bibfnamefont {S.}~\bibnamefont
  {Boixo}}, \bibinfo {author} {\bibfnamefont {S.~V.}\ \bibnamefont {Isakov}},
  \bibinfo {author} {\bibfnamefont {V.~N.}\ \bibnamefont {Smelyanskiy}},
  \bibinfo {author} {\bibfnamefont {R.}~\bibnamefont {Babbush}}, \bibinfo
  {author} {\bibfnamefont {N.}~\bibnamefont {Ding}}, \bibinfo {author}
  {\bibfnamefont {Z.}~\bibnamefont {Jiang}}, \bibinfo {author} {\bibfnamefont
  {M.~J.}\ \bibnamefont {Bremner}}, \bibinfo {author} {\bibfnamefont {J.~M.}\
  \bibnamefont {Martinis}},\ and\ \bibinfo {author} {\bibfnamefont
  {H.}~\bibnamefont {Neven}},\ }\bibfield  {title} {\bibinfo {title}
  {Characterizing quantum supremacy in near-term devices},\ }\href
  {https://doi.org/10.1038/s41567-018-0124-x} {\bibfield  {journal} {\bibinfo
  {journal} {Nature Phys.}\ }\textbf {\bibinfo {volume} {14}},\ \bibinfo
  {pages} {595} (\bibinfo {year} {2018})}\BibitemShut {NoStop}%
\bibitem [{\citenamefont {Solfanelli}\ \emph {et~al.}(2021)\citenamefont
  {Solfanelli}, \citenamefont {Santini},\ and\ \citenamefont
  {Campisi}}]{Solfanelli}%
  \BibitemOpen
  \bibfield  {author} {\bibinfo {author} {\bibfnamefont {A.}~\bibnamefont
  {Solfanelli}}, \bibinfo {author} {\bibfnamefont {A.}~\bibnamefont
  {Santini}},\ and\ \bibinfo {author} {\bibfnamefont {M.}~\bibnamefont
  {Campisi}},\ }\bibfield  {title} {\bibinfo {title} {Experimental verification
  of fluctuation relations with a quantum computer},\ }\href
  {https://doi.org/10.1103/PRXQuantum.2.030353} {\bibfield  {journal} {\bibinfo
   {journal} {PRX Quantum}\ }\textbf {\bibinfo {volume} {2}},\ \bibinfo {pages}
  {030353} (\bibinfo {year} {2021})}\BibitemShut {NoStop}%
\bibitem [{\citenamefont {Hosten}(2022)}]{HostenPhysRevResearch2022}%
  \BibitemOpen
  \bibfield  {author} {\bibinfo {author} {\bibfnamefont {O.}~\bibnamefont
  {Hosten}},\ }\bibfield  {title} {\bibinfo {title} {Constraints on probing
  quantum coherence to infer gravitational entanglement},\ }\href
  {https://doi.org/10.1103/PhysRevResearch.4.013023} {\bibfield  {journal}
  {\bibinfo  {journal} {Phys. Rev. Research}\ }\textbf {\bibinfo {volume}
  {4}},\ \bibinfo {pages} {013023} (\bibinfo {year} {2022})}\BibitemShut
  {NoStop}%
\bibitem [{\citenamefont {Lostaglio}\ \emph {et~al.}(2015)\citenamefont
  {Lostaglio}, \citenamefont {Jennings},\ and\ \citenamefont
  {Rudolph}}]{Lostaglio2015}%
  \BibitemOpen
  \bibfield  {author} {\bibinfo {author} {\bibfnamefont {M.}~\bibnamefont
  {Lostaglio}}, \bibinfo {author} {\bibfnamefont {D.}~\bibnamefont
  {Jennings}},\ and\ \bibinfo {author} {\bibfnamefont {T.}~\bibnamefont
  {Rudolph}},\ }\bibfield  {title} {\bibinfo {title} {Description of quantum
  coherence in thermodynamic processes requires constraints beyond free
  energy},\ }\href {https://doi.org/10.1038/ncomms7383} {\bibfield  {journal}
  {\bibinfo  {journal} {Nature Communications}\ }\textbf {\bibinfo {volume}
  {6}},\ \bibinfo {pages} {6383} (\bibinfo {year} {2015})}\BibitemShut
  {NoStop}%
\bibitem [{\citenamefont {Narasimhachar}\ and\ \citenamefont
  {Gour}(2015)}]{Narasimhachar2015}%
  \BibitemOpen
  \bibfield  {author} {\bibinfo {author} {\bibfnamefont {V.}~\bibnamefont
  {Narasimhachar}}\ and\ \bibinfo {author} {\bibfnamefont {G.}~\bibnamefont
  {Gour}},\ }\href {https://doi.org/10.1038/ncomms8689} {\bibfield  {journal}
  {\bibinfo  {journal} {Nature Communications}\ }\textbf {\bibinfo {volume}
  {6}},\ \bibinfo {pages} {7689} (\bibinfo {year} {2015})}\BibitemShut
  {NoStop}%
\bibitem [{\citenamefont {Klatzow}\ \emph {et~al.}(2019)\citenamefont
  {Klatzow}, \citenamefont {Becker}, \citenamefont {Ledingham}, \citenamefont
  {Weinzetl}, \citenamefont {Kaczmarek}, \citenamefont {Saunders},
  \citenamefont {Nunn}, \citenamefont {Walmsley}, \citenamefont {Uzdin},\ and\
  \citenamefont {Poem}}]{KlatzowPRL2019}%
  \BibitemOpen
  \bibfield  {author} {\bibinfo {author} {\bibfnamefont {J.}~\bibnamefont
  {Klatzow}}, \bibinfo {author} {\bibfnamefont {J.~N.}\ \bibnamefont {Becker}},
  \bibinfo {author} {\bibfnamefont {P.~M.}\ \bibnamefont {Ledingham}}, \bibinfo
  {author} {\bibfnamefont {C.}~\bibnamefont {Weinzetl}}, \bibinfo {author}
  {\bibfnamefont {K.~T.}\ \bibnamefont {Kaczmarek}}, \bibinfo {author}
  {\bibfnamefont {D.~J.}\ \bibnamefont {Saunders}}, \bibinfo {author}
  {\bibfnamefont {J.}~\bibnamefont {Nunn}}, \bibinfo {author} {\bibfnamefont
  {I.~A.}\ \bibnamefont {Walmsley}}, \bibinfo {author} {\bibfnamefont
  {R.}~\bibnamefont {Uzdin}},\ and\ \bibinfo {author} {\bibfnamefont
  {E.}~\bibnamefont {Poem}},\ }\bibfield  {title} {\bibinfo {title}
  {Experimental demonstration of quantum effects in the operation of
  microscopic heat engines},\ }\href
  {https://doi.org/10.1103/PhysRevLett.122.110601} {\bibfield  {journal}
  {\bibinfo  {journal} {Phys. Rev. Lett.}\ }\textbf {\bibinfo {volume} {122}},\
  \bibinfo {pages} {110601} (\bibinfo {year} {2019})}\BibitemShut {NoStop}%
\bibitem [{\citenamefont {Trushechkin}\ \emph {et~al.}(2022)\citenamefont
  {Trushechkin}, \citenamefont {Merkli}, \citenamefont {Cresser},\ and\
  \citenamefont {Anders}}]{TrushechkinQuSci2022}%
  \BibitemOpen
  \bibfield  {author} {\bibinfo {author} {\bibfnamefont {A.~S.}\ \bibnamefont
  {Trushechkin}}, \bibinfo {author} {\bibfnamefont {M.}~\bibnamefont {Merkli}},
  \bibinfo {author} {\bibfnamefont {J.~D.}\ \bibnamefont {Cresser}},\ and\
  \bibinfo {author} {\bibfnamefont {J.}~\bibnamefont {Anders}},\ }\bibfield
  {title} {\bibinfo {title} {Open quantum system dynamics and the mean force
  gibbs state},\ }\href {https://doi.org/10.1116/5.0073853} {\bibfield
  {journal} {\bibinfo  {journal} {AVS Quantum Science}\ }\textbf {\bibinfo
  {volume} {4}},\ \bibinfo {pages} {012301} (\bibinfo {year}
  {2022})}\BibitemShut {NoStop}%
\bibitem [{\citenamefont {Linden}\ \emph {et~al.}(2009)\citenamefont {Linden},
  \citenamefont {Popescu}, \citenamefont {Short},\ and\ \citenamefont
  {Winter}}]{Linden2009}%
  \BibitemOpen
  \bibfield  {author} {\bibinfo {author} {\bibfnamefont {N.}~\bibnamefont
  {Linden}}, \bibinfo {author} {\bibfnamefont {S.}~\bibnamefont {Popescu}},
  \bibinfo {author} {\bibfnamefont {A.~J.}\ \bibnamefont {Short}},\ and\
  \bibinfo {author} {\bibfnamefont {A.}~\bibnamefont {Winter}},\ }\bibfield
  {title} {\bibinfo {title} {Quantum mechanical evolution towards thermal
  equilibrium},\ }\bibfield  {journal} {\bibinfo  {journal} {Physical Review
  E}\ }\textbf {\bibinfo {volume} {79}},\ \href
  {https://doi.org/10.1103/physreve.79.061103} {10.1103/physreve.79.061103}
  (\bibinfo {year} {2009})\BibitemShut {NoStop}%
\bibitem [{\citenamefont {Farrelly}\ \emph {et~al.}(2017)\citenamefont
  {Farrelly}, \citenamefont {Brand{\~{a}}o},\ and\ \citenamefont
  {Cramer}}]{Farrelly2017}%
  \BibitemOpen
  \bibfield  {author} {\bibinfo {author} {\bibfnamefont {T.}~\bibnamefont
  {Farrelly}}, \bibinfo {author} {\bibfnamefont {F.~G.}\ \bibnamefont
  {Brand{\~{a}}o}},\ and\ \bibinfo {author} {\bibfnamefont {M.}~\bibnamefont
  {Cramer}},\ }\bibfield  {title} {\bibinfo {title} {Thermalization and return
  to equilibrium on finite quantum lattice systems},\ }\bibfield  {journal}
  {\bibinfo  {journal} {Physical Review Letters}\ }\textbf {\bibinfo {volume}
  {118}},\ \href {https://doi.org/10.1103/physrevlett.118.140601}
  {10.1103/physrevlett.118.140601} (\bibinfo {year} {2017})\BibitemShut
  {NoStop}%
\bibitem [{\citenamefont {Uzdin}\ and\ \citenamefont
  {Rahav}(2018)}]{UzdinPRX2018}%
  \BibitemOpen
  \bibfield  {author} {\bibinfo {author} {\bibfnamefont {R.}~\bibnamefont
  {Uzdin}}\ and\ \bibinfo {author} {\bibfnamefont {S.}~\bibnamefont {Rahav}},\
  }\bibfield  {title} {\bibinfo {title} {Global passivity in microscopic
  thermodynamics},\ }\href {https://doi.org/10.1103/PhysRevX.8.021064}
  {\bibfield  {journal} {\bibinfo  {journal} {Phys. Rev. X}\ }\textbf {\bibinfo
  {volume} {8}},\ \bibinfo {pages} {021064} (\bibinfo {year}
  {2018})}\BibitemShut {NoStop}%
\bibitem [{\citenamefont {Breuer}\ \emph {et~al.}(2002)\citenamefont {Breuer},
  \citenamefont {Petruccione} \emph {et~al.}}]{breuer2002theory}%
  \BibitemOpen
  \bibfield  {author} {\bibinfo {author} {\bibfnamefont {H.-P.}\ \bibnamefont
  {Breuer}}, \bibinfo {author} {\bibfnamefont {F.}~\bibnamefont {Petruccione}},
  \emph {et~al.},\ }\href@noop {} {\emph {\bibinfo {title} {The theory of open
  quantum systems}}}\ (\bibinfo  {publisher} {Oxford University Press on
  Demand},\ \bibinfo {year} {2002})\BibitemShut {NoStop}%
\bibitem [{\citenamefont {Talkner}\ and\ \citenamefont
  {H\"anggi}(2016)}]{Hanggi}%
  \BibitemOpen
  \bibfield  {author} {\bibinfo {author} {\bibfnamefont {P.}~\bibnamefont
  {Talkner}}\ and\ \bibinfo {author} {\bibfnamefont {P.}~\bibnamefont
  {H\"anggi}},\ }\bibfield  {title} {\bibinfo {title} {Open system trajectories
  specify fluctuating work but not heat},\ }\href
  {https://doi.org/10.1103/PhysRevE.94.022143} {\bibfield  {journal} {\bibinfo
  {journal} {Phys. Rev. E}\ }\textbf {\bibinfo {volume} {94}},\ \bibinfo
  {pages} {022143} (\bibinfo {year} {2016})}\BibitemShut {NoStop}%
\bibitem [{\citenamefont {Kubo}\ \emph {et~al.}(1985)\citenamefont {Kubo},
  \citenamefont {Toda},\ and\ \citenamefont {Hashitsume}}]{Kubo1985}%
  \BibitemOpen
  \bibfield  {author} {\bibinfo {author} {\bibfnamefont {R.}~\bibnamefont
  {Kubo}}, \bibinfo {author} {\bibfnamefont {M.}~\bibnamefont {Toda}},\ and\
  \bibinfo {author} {\bibfnamefont {N.}~\bibnamefont {Hashitsume}},\ }\href
  {https://doi.org/10.1007/978-3-642-96701-6} {\emph {\bibinfo {title}
  {Statistical Physics {II}}}}\ (\bibinfo  {publisher} {Springer Berlin
  Heidelberg},\ \bibinfo {year} {1985})\BibitemShut {NoStop}%
\bibitem [{\citenamefont {Mori}\ and\ \citenamefont
  {Miyashita}(2008)}]{Mori2008}%
  \BibitemOpen
  \bibfield  {author} {\bibinfo {author} {\bibfnamefont {T.}~\bibnamefont
  {Mori}}\ and\ \bibinfo {author} {\bibfnamefont {S.}~\bibnamefont
  {Miyashita}},\ }\bibfield  {title} {\bibinfo {title} {Dynamics of the density
  matrix in contact with a thermal bath and the quantum master equation},\
  }\href {https://doi.org/10.1143/jpsj.77.124005} {\bibfield  {journal}
  {\bibinfo  {journal} {Journal of the Physical Society of Japan}\ }\textbf
  {\bibinfo {volume} {77}},\ \bibinfo {pages} {124005} (\bibinfo {year}
  {2008})}\BibitemShut {NoStop}%
\bibitem [{\citenamefont {Walls}(1970)}]{Walls1970}%
  \BibitemOpen
  \bibfield  {author} {\bibinfo {author} {\bibfnamefont {D.~F.}\ \bibnamefont
  {Walls}},\ }\bibfield  {title} {\bibinfo {title} {Higher order effects in the
  master equation for coupled systems},\ }\href
  {https://doi.org/10.1007/BF01396784} {\bibfield  {journal} {\bibinfo
  {journal} {Zeitschrift für Physik A Hadrons and nuclei}\ }\textbf {\bibinfo
  {volume} {234}},\ \bibinfo {pages} {231} (\bibinfo {year}
  {1970})}\BibitemShut {NoStop}%
\bibitem [{\citenamefont {Carmichael}\ and\ \citenamefont
  {Walls}(1973)}]{Carmichael_1973}%
  \BibitemOpen
  \bibfield  {author} {\bibinfo {author} {\bibfnamefont {H.~J.}\ \bibnamefont
  {Carmichael}}\ and\ \bibinfo {author} {\bibfnamefont {D.~F.}\ \bibnamefont
  {Walls}},\ }\bibfield  {title} {\bibinfo {title} {Master equation for
  strongly interacting systems},\ }\href
  {https://doi.org/10.1088/0305-4470/6/10/014} {\bibfield  {journal} {\bibinfo
  {journal} {Journal of Physics A: Mathematical, Nuclear and General}\ }\textbf
  {\bibinfo {volume} {6}},\ \bibinfo {pages} {1552} (\bibinfo {year}
  {1973})}\BibitemShut {NoStop}%
\bibitem [{\citenamefont {Cattaneo}\ \emph {et~al.}(2019)\citenamefont
  {Cattaneo}, \citenamefont {Giorgi}, \citenamefont {Maniscalco},\ and\
  \citenamefont {Zambrini}}]{Cattaneo_2019}%
  \BibitemOpen
  \bibfield  {author} {\bibinfo {author} {\bibfnamefont {M.}~\bibnamefont
  {Cattaneo}}, \bibinfo {author} {\bibfnamefont {G.~L.}\ \bibnamefont
  {Giorgi}}, \bibinfo {author} {\bibfnamefont {S.}~\bibnamefont {Maniscalco}},\
  and\ \bibinfo {author} {\bibfnamefont {R.}~\bibnamefont {Zambrini}},\
  }\bibfield  {title} {\bibinfo {title} {Local versus global master equation
  with common and separate baths: superiority of the global approach in partial
  secular approximation},\ }\href {https://doi.org/10.1088/1367-2630/ab54ac}
  {\bibfield  {journal} {\bibinfo  {journal} {New Journal of Physics}\ }\textbf
  {\bibinfo {volume} {21}},\ \bibinfo {pages} {113045} (\bibinfo {year}
  {2019})}\BibitemShut {NoStop}%
\bibitem [{\citenamefont {Trushechkin}(2021)}]{TrushechkinPRA2021}%
  \BibitemOpen
  \bibfield  {author} {\bibinfo {author} {\bibfnamefont {A.}~\bibnamefont
  {Trushechkin}},\ }\bibfield  {title} {\bibinfo {title} {Unified
  gorini-kossakowski-lindblad-sudarshan quantum master equation beyond the
  secular approximation},\ }\href {https://doi.org/10.1103/PhysRevA.103.062226}
  {\bibfield  {journal} {\bibinfo  {journal} {Phys. Rev. A}\ }\textbf {\bibinfo
  {volume} {103}},\ \bibinfo {pages} {062226} (\bibinfo {year}
  {2021})}\BibitemShut {NoStop}%
\bibitem [{\citenamefont {Ronzani}\ \emph {et~al.}(2018)\citenamefont
  {Ronzani}, \citenamefont {Karimi}, \citenamefont {Senior}, \citenamefont
  {Chang}, \citenamefont {Peltonen}, \citenamefont {Chen},\ and\ \citenamefont
  {Pekola}}]{ronzaniNature2018}%
  \BibitemOpen
  \bibfield  {author} {\bibinfo {author} {\bibfnamefont {A.}~\bibnamefont
  {Ronzani}}, \bibinfo {author} {\bibfnamefont {B.}~\bibnamefont {Karimi}},
  \bibinfo {author} {\bibfnamefont {J.}~\bibnamefont {Senior}}, \bibinfo
  {author} {\bibfnamefont {Y.-C.}\ \bibnamefont {Chang}}, \bibinfo {author}
  {\bibfnamefont {J.~T.}\ \bibnamefont {Peltonen}}, \bibinfo {author}
  {\bibfnamefont {C.}~\bibnamefont {Chen}},\ and\ \bibinfo {author}
  {\bibfnamefont {J.~P.}\ \bibnamefont {Pekola}},\ }\bibfield  {title}
  {\bibinfo {title} {Tunable photonic heat transport in a quantum heat valve},\
  }\href {https://doi.org/https://doi.org/10.1038/s41567-018-0199-4} {\bibfield
   {journal} {\bibinfo  {journal} {Nature Physics}\ }\textbf {\bibinfo {volume}
  {14}},\ \bibinfo {pages} {991} (\bibinfo {year} {2018})}\BibitemShut
  {NoStop}%
\bibitem [{\citenamefont {Pekola}\ and\ \citenamefont
  {Karimi}(2021)}]{PekolaRevModPhys2021}%
  \BibitemOpen
  \bibfield  {author} {\bibinfo {author} {\bibfnamefont {J.~P.}\ \bibnamefont
  {Pekola}}\ and\ \bibinfo {author} {\bibfnamefont {B.}~\bibnamefont
  {Karimi}},\ }\bibfield  {title} {\bibinfo {title} {Colloquium: Quantum heat
  transport in condensed matter systems},\ }\href
  {https://doi.org/10.1103/RevModPhys.93.041001} {\bibfield  {journal}
  {\bibinfo  {journal} {Rev. Mod. Phys.}\ }\textbf {\bibinfo {volume} {93}},\
  \bibinfo {pages} {041001} (\bibinfo {year} {2021})}\BibitemShut {NoStop}%
\bibitem [{\citenamefont {Konopik}\ and\ \citenamefont
  {Lutz}(2022)}]{KonopikPRRes2022}%
  \BibitemOpen
  \bibfield  {author} {\bibinfo {author} {\bibfnamefont {M.}~\bibnamefont
  {Konopik}}\ and\ \bibinfo {author} {\bibfnamefont {E.}~\bibnamefont {Lutz}},\
  }\bibfield  {title} {\bibinfo {title} {Local master equations may fail to
  describe dissipative critical behavior},\ }\href
  {https://doi.org/10.1103/PhysRevResearch.4.013171} {\bibfield  {journal}
  {\bibinfo  {journal} {Phys. Rev. Research}\ }\textbf {\bibinfo {volume}
  {4}},\ \bibinfo {pages} {013171} (\bibinfo {year} {2022})}\BibitemShut
  {NoStop}%
\bibitem [{\citenamefont {Guarnieri}\ \emph {et~al.}(2018)\citenamefont
  {Guarnieri}, \citenamefont {Kol\'a\ifmmode~\check{r}\else \v{r}\fi{}},\ and\
  \citenamefont {Filip}}]{giacomoPRL2018}%
  \BibitemOpen
  \bibfield  {author} {\bibinfo {author} {\bibfnamefont {G.}~\bibnamefont
  {Guarnieri}}, \bibinfo {author} {\bibfnamefont {M.}~\bibnamefont
  {Kol\'a\ifmmode~\check{r}\else \v{r}\fi{}}},\ and\ \bibinfo {author}
  {\bibfnamefont {R.}~\bibnamefont {Filip}},\ }\bibfield  {title} {\bibinfo
  {title} {Steady-state coherences by composite system-bath interactions},\
  }\href {https://doi.org/https://doi.org/10.1103/PhysRevLett.121.070401}
  {\bibfield  {journal} {\bibinfo  {journal} {Phys. Rev. Lett.}\ }\textbf
  {\bibinfo {volume} {121}},\ \bibinfo {pages} {070401} (\bibinfo {year}
  {2018})}\BibitemShut {NoStop}%
\bibitem [{\citenamefont {Guarnieri}\ \emph {et~al.}(2020)\citenamefont
  {Guarnieri}, \citenamefont {Morrone}, \citenamefont {Çakmak}, \citenamefont
  {Plastina},\ and\ \citenamefont {Campbell}}]{GUARNIERIPLA2020}%
  \BibitemOpen
  \bibfield  {author} {\bibinfo {author} {\bibfnamefont {G.}~\bibnamefont
  {Guarnieri}}, \bibinfo {author} {\bibfnamefont {D.}~\bibnamefont {Morrone}},
  \bibinfo {author} {\bibfnamefont {B.}~\bibnamefont {Çakmak}}, \bibinfo
  {author} {\bibfnamefont {F.}~\bibnamefont {Plastina}},\ and\ \bibinfo
  {author} {\bibfnamefont {S.}~\bibnamefont {Campbell}},\ }\bibfield  {title}
  {\bibinfo {title} {Non-equilibrium steady-states of memoryless quantum
  collision models},\ }\href
  {https://doi.org/https://doi.org/10.1016/j.physleta.2020.126576} {\bibfield
  {journal} {\bibinfo  {journal} {Physics Letters A}\ }\textbf {\bibinfo
  {volume} {384}},\ \bibinfo {pages} {126576} (\bibinfo {year}
  {2020})}\BibitemShut {NoStop}%
\bibitem [{\citenamefont {Purkayastha}\ \emph {et~al.}(2020)\citenamefont
  {Purkayastha}, \citenamefont {Guarnieri}, \citenamefont {Mitchison},
  \citenamefont {Filip},\ and\ \citenamefont {Goold}}]{purkayastha2020tunable}%
  \BibitemOpen
  \bibfield  {author} {\bibinfo {author} {\bibfnamefont {A.}~\bibnamefont
  {Purkayastha}}, \bibinfo {author} {\bibfnamefont {G.}~\bibnamefont
  {Guarnieri}}, \bibinfo {author} {\bibfnamefont {M.~T.}\ \bibnamefont
  {Mitchison}}, \bibinfo {author} {\bibfnamefont {R.}~\bibnamefont {Filip}},\
  and\ \bibinfo {author} {\bibfnamefont {J.}~\bibnamefont {Goold}},\ }\bibfield
   {title} {\bibinfo {title} {Tunable phonon-induced steady-state coherence in
  a double-quantum-dot charge qubit},\ }\href
  {https://doi.org/https://doi.org/10.1038/s41534-020-0256-6} {\bibfield
  {journal} {\bibinfo  {journal} {npj Quantum Information}\ }\textbf {\bibinfo
  {volume} {6}},\ \bibinfo {pages} {1} (\bibinfo {year} {2020})}\BibitemShut
  {NoStop}%
\bibitem [{\citenamefont {Cresser}\ and\ \citenamefont
  {Anders}(2021)}]{AndersPRL2021strongGibbs}%
  \BibitemOpen
  \bibfield  {author} {\bibinfo {author} {\bibfnamefont {J.~D.}\ \bibnamefont
  {Cresser}}\ and\ \bibinfo {author} {\bibfnamefont {J.}~\bibnamefont
  {Anders}},\ }\bibfield  {title} {\bibinfo {title} {Weak and ultrastrong
  coupling limits of the quantum mean force gibbs state},\ }\href
  {https://doi.org/10.1103/PhysRevLett.127.250601} {\bibfield  {journal}
  {\bibinfo  {journal} {Phys. Rev. Lett.}\ }\textbf {\bibinfo {volume} {127}},\
  \bibinfo {pages} {250601} (\bibinfo {year} {2021})}\BibitemShut {NoStop}%
\bibitem [{\citenamefont {Slobodeniuk}\ \emph {et~al.}(2022)\citenamefont
  {Slobodeniuk}, \citenamefont {Novotn{\'{y}}},\ and\ \citenamefont
  {Filip}}]{slobodeniuk2021extraction}%
  \BibitemOpen
  \bibfield  {author} {\bibinfo {author} {\bibfnamefont {A.}~\bibnamefont
  {Slobodeniuk}}, \bibinfo {author} {\bibfnamefont {T.}~\bibnamefont
  {Novotn{\'{y}}}},\ and\ \bibinfo {author} {\bibfnamefont {R.}~\bibnamefont
  {Filip}},\ }\bibfield  {title} {\bibinfo {title} {Extraction of autonomous
  quantum coherences},\ }\href
  {https://doi.org/https://doi.org/10.22331/q-2022-04-15-689} {\bibfield
  {journal} {\bibinfo  {journal} {{Quantum}}\ }\textbf {\bibinfo {volume}
  {6}},\ \bibinfo {pages} {689} (\bibinfo {year} {2022})}\BibitemShut {NoStop}%
\bibitem [{\citenamefont {Rom\'an-Ancheyta}\ \emph {et~al.}(2021)\citenamefont
  {Rom\'an-Ancheyta}, \citenamefont {Kol\'a\ifmmode~\check{r}\else \v{r}\fi{}},
  \citenamefont {Guarnieri},\ and\ \citenamefont {Filip}}]{RicardoPRA2021}%
  \BibitemOpen
  \bibfield  {author} {\bibinfo {author} {\bibfnamefont {R.}~\bibnamefont
  {Rom\'an-Ancheyta}}, \bibinfo {author} {\bibfnamefont {M.}~\bibnamefont
  {Kol\'a\ifmmode~\check{r}\else \v{r}\fi{}}}, \bibinfo {author} {\bibfnamefont
  {G.}~\bibnamefont {Guarnieri}},\ and\ \bibinfo {author} {\bibfnamefont
  {R.}~\bibnamefont {Filip}},\ }\bibfield  {title} {\bibinfo {title} {Enhanced
  steady-state coherence via repeated system-bath interactions},\ }\href
  {https://doi.org/10.1103/PhysRevA.104.062209} {\bibfield  {journal} {\bibinfo
   {journal} {Phys. Rev. A}\ }\textbf {\bibinfo {volume} {104}},\ \bibinfo
  {pages} {062209} (\bibinfo {year} {2021})}\BibitemShut {NoStop}%
\bibitem [{\citenamefont {Birrittella}\ \emph {et~al.}(2021)\citenamefont
  {Birrittella}, \citenamefont {Alsing},\ and\ \citenamefont
  {Gerry}}]{AVSBirrittella}%
  \BibitemOpen
  \bibfield  {author} {\bibinfo {author} {\bibfnamefont {R.~J.}\ \bibnamefont
  {Birrittella}}, \bibinfo {author} {\bibfnamefont {P.~M.}\ \bibnamefont
  {Alsing}},\ and\ \bibinfo {author} {\bibfnamefont {C.~C.}\ \bibnamefont
  {Gerry}},\ }\bibfield  {title} {\bibinfo {title} {The parity operator:
  Applications in quantum metrology},\ }\href
  {https://doi.org/10.1116/5.0026148} {\bibfield  {journal} {\bibinfo
  {journal} {AVS Quantum Science}\ }\textbf {\bibinfo {volume} {3}},\ \bibinfo
  {pages} {014701} (\bibinfo {year} {2021})},\ \Eprint
  {https://arxiv.org/abs/https://doi.org/10.1116/5.0026148}
  {https://doi.org/10.1116/5.0026148} \BibitemShut {NoStop}%
\bibitem [{\citenamefont {Braunstein}\ and\ \citenamefont
  {Caves}(1994)}]{BraunsteinPRL1994}%
  \BibitemOpen
  \bibfield  {author} {\bibinfo {author} {\bibfnamefont {S.~L.}\ \bibnamefont
  {Braunstein}}\ and\ \bibinfo {author} {\bibfnamefont {C.~M.}\ \bibnamefont
  {Caves}},\ }\bibfield  {title} {\bibinfo {title} {Statistical distance and
  the geometry of quantum states},\ }\href
  {https://doi.org/10.1103/PhysRevLett.72.3439} {\bibfield  {journal} {\bibinfo
   {journal} {Phys. Rev. Lett.}\ }\textbf {\bibinfo {volume} {72}},\ \bibinfo
  {pages} {3439} (\bibinfo {year} {1994})}\BibitemShut {NoStop}%
\bibitem [{\citenamefont {Baumgratz}\ \emph {et~al.}(2014)\citenamefont
  {Baumgratz}, \citenamefont {Cramer},\ and\ \citenamefont
  {Plenio}}]{baumgratz2014}%
  \BibitemOpen
  \bibfield  {author} {\bibinfo {author} {\bibfnamefont {T.}~\bibnamefont
  {Baumgratz}}, \bibinfo {author} {\bibfnamefont {M.}~\bibnamefont {Cramer}},\
  and\ \bibinfo {author} {\bibfnamefont {M.~B.}\ \bibnamefont {Plenio}},\
  }\bibfield  {title} {\bibinfo {title} {Quantifying coherence},\ }\href
  {https://doi.org/https://doi.org/10.1103/PhysRevLett.113.140401} {\bibfield
  {journal} {\bibinfo  {journal} {Phys. Rev. Lett.}\ }\textbf {\bibinfo
  {volume} {113}},\ \bibinfo {pages} {140401} (\bibinfo {year}
  {2014})}\BibitemShut {NoStop}%
\bibitem [{\citenamefont {Sylju\aa{}sen}(2003)}]{OlavPRA2003}%
  \BibitemOpen
  \bibfield  {author} {\bibinfo {author} {\bibfnamefont {O.~F.}\ \bibnamefont
  {Sylju\aa{}sen}},\ }\bibfield  {title} {\bibinfo {title} {Entanglement and
  spontaneous symmetry breaking in quantum spin models},\ }\href
  {https://doi.org/10.1103/PhysRevA.68.060301} {\bibfield  {journal} {\bibinfo
  {journal} {Phys. Rev. A}\ }\textbf {\bibinfo {volume} {68}},\ \bibinfo
  {pages} {060301} (\bibinfo {year} {2003})}\BibitemShut {NoStop}%
\bibitem [{\citenamefont {Justino}\ and\ \citenamefont
  {de~Oliveira}(2012)}]{JustinoPRA2012chainsymmetries}%
  \BibitemOpen
  \bibfield  {author} {\bibinfo {author} {\bibfnamefont {L.}~\bibnamefont
  {Justino}}\ and\ \bibinfo {author} {\bibfnamefont {T.~R.}\ \bibnamefont
  {de~Oliveira}},\ }\bibfield  {title} {\bibinfo {title} {Bell inequalities and
  entanglement at quantum phase transitions in the $\mathit{XXZ}$ model},\
  }\href {https://doi.org/10.1103/PhysRevA.85.052128} {\bibfield  {journal}
  {\bibinfo  {journal} {Phys. Rev. A}\ }\textbf {\bibinfo {volume} {85}},\
  \bibinfo {pages} {052128} (\bibinfo {year} {2012})}\BibitemShut {NoStop}%
\bibitem [{\citenamefont {Gu}\ \emph {et~al.}(2017)\citenamefont {Gu},
  \citenamefont {Kockum}, \citenamefont {Miranowicz}, \citenamefont {xi~Liu},\
  and\ \citenamefont {Nori}}]{GUPhysRep2017}%
  \BibitemOpen
  \bibfield  {author} {\bibinfo {author} {\bibfnamefont {X.}~\bibnamefont
  {Gu}}, \bibinfo {author} {\bibfnamefont {A.~F.}\ \bibnamefont {Kockum}},
  \bibinfo {author} {\bibfnamefont {A.}~\bibnamefont {Miranowicz}}, \bibinfo
  {author} {\bibfnamefont {Y.}~\bibnamefont {xi~Liu}},\ and\ \bibinfo {author}
  {\bibfnamefont {F.}~\bibnamefont {Nori}},\ }\bibfield  {title} {\bibinfo
  {title} {Microwave photonics with superconducting quantum circuits},\ }\href
  {https://doi.org/https://doi.org/10.1016/j.physrep.2017.10.002} {\bibfield
  {journal} {\bibinfo  {journal} {Physics Reports}\ }\textbf {\bibinfo {volume}
  {718-719}},\ \bibinfo {pages} {1} (\bibinfo {year} {2017})},\ \bibinfo {note}
  {microwave photonics with superconducting quantum circuits}\BibitemShut
  {NoStop}%
\bibitem [{\citenamefont {Blais}\ \emph {et~al.}(2004)\citenamefont {Blais},
  \citenamefont {Huang}, \citenamefont {Wallraff}, \citenamefont {Girvin},\
  and\ \citenamefont {Schoelkopf}}]{BlaisPRA2004Transmon1}%
  \BibitemOpen
  \bibfield  {author} {\bibinfo {author} {\bibfnamefont {A.}~\bibnamefont
  {Blais}}, \bibinfo {author} {\bibfnamefont {R.-S.}\ \bibnamefont {Huang}},
  \bibinfo {author} {\bibfnamefont {A.}~\bibnamefont {Wallraff}}, \bibinfo
  {author} {\bibfnamefont {S.~M.}\ \bibnamefont {Girvin}},\ and\ \bibinfo
  {author} {\bibfnamefont {R.~J.}\ \bibnamefont {Schoelkopf}},\ }\bibfield
  {title} {\bibinfo {title} {Cavity quantum electrodynamics for superconducting
  electrical circuits: An architecture for quantum computation},\ }\href
  {https://doi.org/10.1103/PhysRevA.69.062320} {\bibfield  {journal} {\bibinfo
  {journal} {Phys. Rev. A}\ }\textbf {\bibinfo {volume} {69}},\ \bibinfo
  {pages} {062320} (\bibinfo {year} {2004})}\BibitemShut {NoStop}%
\bibitem [{\citenamefont {Koch}\ \emph {et~al.}(2007)\citenamefont {Koch},
  \citenamefont {Yu}, \citenamefont {Gambetta}, \citenamefont {Houck},
  \citenamefont {Schuster}, \citenamefont {Majer}, \citenamefont {Blais},
  \citenamefont {Devoret}, \citenamefont {Girvin},\ and\ \citenamefont
  {Schoelkopf}}]{KochPRA2007Transmon2}%
  \BibitemOpen
  \bibfield  {author} {\bibinfo {author} {\bibfnamefont {J.}~\bibnamefont
  {Koch}}, \bibinfo {author} {\bibfnamefont {T.~M.}\ \bibnamefont {Yu}},
  \bibinfo {author} {\bibfnamefont {J.}~\bibnamefont {Gambetta}}, \bibinfo
  {author} {\bibfnamefont {A.~A.}\ \bibnamefont {Houck}}, \bibinfo {author}
  {\bibfnamefont {D.~I.}\ \bibnamefont {Schuster}}, \bibinfo {author}
  {\bibfnamefont {J.}~\bibnamefont {Majer}}, \bibinfo {author} {\bibfnamefont
  {A.}~\bibnamefont {Blais}}, \bibinfo {author} {\bibfnamefont {M.~H.}\
  \bibnamefont {Devoret}}, \bibinfo {author} {\bibfnamefont {S.~M.}\
  \bibnamefont {Girvin}},\ and\ \bibinfo {author} {\bibfnamefont {R.~J.}\
  \bibnamefont {Schoelkopf}},\ }\bibfield  {title} {\bibinfo {title}
  {Charge-insensitive qubit design derived from the cooper pair box},\ }\href
  {https://doi.org/10.1103/PhysRevA.76.042319} {\bibfield  {journal} {\bibinfo
  {journal} {Phys. Rev. A}\ }\textbf {\bibinfo {volume} {76}},\ \bibinfo
  {pages} {042319} (\bibinfo {year} {2007})}\BibitemShut {NoStop}%
\bibitem [{\citenamefont {Raja}\ \emph {et~al.}(2021)\citenamefont {Raja},
  \citenamefont {Maniscalco}, \citenamefont {Paraoanu}, \citenamefont
  {Pekola},\ and\ \citenamefont {Gullo}}]{Hamedani_Pekola_Entropy2021}%
  \BibitemOpen
  \bibfield  {author} {\bibinfo {author} {\bibfnamefont {S.~H.}\ \bibnamefont
  {Raja}}, \bibinfo {author} {\bibfnamefont {S.}~\bibnamefont {Maniscalco}},
  \bibinfo {author} {\bibfnamefont {G.~S.}\ \bibnamefont {Paraoanu}}, \bibinfo
  {author} {\bibfnamefont {J.~P.}\ \bibnamefont {Pekola}},\ and\ \bibinfo
  {author} {\bibfnamefont {N.~L.}\ \bibnamefont {Gullo}},\ }\bibfield  {title}
  {\bibinfo {title} {Finite-time quantum stirling heat engine},\ }\href
  {https://doi.org/10.1088/1367-2630/abe9d7} {\bibfield  {journal} {\bibinfo
  {journal} {New Journal of Physics}\ }\textbf {\bibinfo {volume} {23}},\
  \bibinfo {pages} {033034} (\bibinfo {year} {2021})}\BibitemShut {NoStop}%
\bibitem [{\citenamefont {Forn-D\'{\i}az}\ \emph {et~al.}(2010)\citenamefont
  {Forn-D\'{\i}az}, \citenamefont {Lisenfeld}, \citenamefont {Marcos},
  \citenamefont {Garc\'{\i}a-Ripoll}, \citenamefont {Solano}, \citenamefont
  {Harmans},\ and\ \citenamefont {Mooij}}]{LisenfeldPRL2010}%
  \BibitemOpen
  \bibfield  {author} {\bibinfo {author} {\bibfnamefont {P.}~\bibnamefont
  {Forn-D\'{\i}az}}, \bibinfo {author} {\bibfnamefont {J.}~\bibnamefont
  {Lisenfeld}}, \bibinfo {author} {\bibfnamefont {D.}~\bibnamefont {Marcos}},
  \bibinfo {author} {\bibfnamefont {J.~J.}\ \bibnamefont {Garc\'{\i}a-Ripoll}},
  \bibinfo {author} {\bibfnamefont {E.}~\bibnamefont {Solano}}, \bibinfo
  {author} {\bibfnamefont {C.~J. P.~M.}\ \bibnamefont {Harmans}},\ and\
  \bibinfo {author} {\bibfnamefont {J.~E.}\ \bibnamefont {Mooij}},\ }\bibfield
  {title} {\bibinfo {title} {Observation of the bloch-siegert shift in a
  qubit-oscillator system in the ultrastrong coupling regime},\ }\href
  {https://doi.org/10.1103/PhysRevLett.105.237001} {\bibfield  {journal}
  {\bibinfo  {journal} {Phys. Rev. Lett.}\ }\textbf {\bibinfo {volume} {105}},\
  \bibinfo {pages} {237001} (\bibinfo {year} {2010})}\BibitemShut {NoStop}%
\bibitem [{\citenamefont {Guthrie}\ \emph {et~al.}(2022)\citenamefont
  {Guthrie}, \citenamefont {Satrya}, \citenamefont {Chang}, \citenamefont
  {Menczel}, \citenamefont {Nori},\ and\ \citenamefont
  {Pekola}}]{GuthriePhysRevApp2022}%
  \BibitemOpen
  \bibfield  {author} {\bibinfo {author} {\bibfnamefont {A.}~\bibnamefont
  {Guthrie}}, \bibinfo {author} {\bibfnamefont {C.~D.}\ \bibnamefont {Satrya}},
  \bibinfo {author} {\bibfnamefont {Y.-C.}\ \bibnamefont {Chang}}, \bibinfo
  {author} {\bibfnamefont {P.}~\bibnamefont {Menczel}}, \bibinfo {author}
  {\bibfnamefont {F.}~\bibnamefont {Nori}},\ and\ \bibinfo {author}
  {\bibfnamefont {J.~P.}\ \bibnamefont {Pekola}},\ }\bibfield  {title}
  {\bibinfo {title} {Cooper-pair box coupled to two resonators: An architecture
  for a quantum refrigerator},\ }\href
  {https://doi.org/10.1103/PhysRevApplied.17.064022} {\bibfield  {journal}
  {\bibinfo  {journal} {Phys. Rev. Applied}\ }\textbf {\bibinfo {volume}
  {17}},\ \bibinfo {pages} {064022} (\bibinfo {year} {2022})}\BibitemShut
  {NoStop}%
\bibitem [{\citenamefont {Kol\'a\ifmmode~\check{r}\else \v{r}\fi{}}\ \emph
  {et~al.}(2017)\citenamefont {Kol\'a\ifmmode~\check{r}\else \v{r}\fi{}},
  \citenamefont {Ryabov},\ and\ \citenamefont {Filip}}]{ArtaPRA2017}%
  \BibitemOpen
  \bibfield  {author} {\bibinfo {author} {\bibfnamefont {M.}~\bibnamefont
  {Kol\'a\ifmmode~\check{r}\else \v{r}\fi{}}}, \bibinfo {author} {\bibfnamefont
  {A.}~\bibnamefont {Ryabov}},\ and\ \bibinfo {author} {\bibfnamefont
  {R.}~\bibnamefont {Filip}},\ }\bibfield  {title} {\bibinfo {title}
  {Optomechanical oscillator controlled by variation in its heat bath
  temperature},\ }\href {https://doi.org/10.1103/PhysRevA.95.042105} {\bibfield
   {journal} {\bibinfo  {journal} {Phys. Rev. A}\ }\textbf {\bibinfo {volume}
  {95}},\ \bibinfo {pages} {042105} (\bibinfo {year} {2017})}\BibitemShut
  {NoStop}%
\bibitem [{\citenamefont {Brivio}\ \emph {et~al.}(2010)\citenamefont {Brivio},
  \citenamefont {Cialdi}, \citenamefont {Vezzoli}, \citenamefont {Gebrehiwot},
  \citenamefont {Genoni}, \citenamefont {Olivares},\ and\ \citenamefont
  {Paris}}]{BrivioPRA2010}%
  \BibitemOpen
  \bibfield  {author} {\bibinfo {author} {\bibfnamefont {D.}~\bibnamefont
  {Brivio}}, \bibinfo {author} {\bibfnamefont {S.}~\bibnamefont {Cialdi}},
  \bibinfo {author} {\bibfnamefont {S.}~\bibnamefont {Vezzoli}}, \bibinfo
  {author} {\bibfnamefont {B.~T.}\ \bibnamefont {Gebrehiwot}}, \bibinfo
  {author} {\bibfnamefont {M.~G.}\ \bibnamefont {Genoni}}, \bibinfo {author}
  {\bibfnamefont {S.}~\bibnamefont {Olivares}},\ and\ \bibinfo {author}
  {\bibfnamefont {M.~G.~A.}\ \bibnamefont {Paris}},\ }\bibfield  {title}
  {\bibinfo {title} {Experimental estimation of one-parameter qubit gates in
  the presence of phase diffusion},\ }\href
  {https://doi.org/10.1103/PhysRevA.81.012305} {\bibfield  {journal} {\bibinfo
  {journal} {Phys. Rev. A}\ }\textbf {\bibinfo {volume} {81}},\ \bibinfo
  {pages} {012305} (\bibinfo {year} {2010})}\BibitemShut {NoStop}%
\bibitem [{\citenamefont {Preskill}(2018)}]{Preskill2018quantumcomputing}%
  \BibitemOpen
  \bibfield  {author} {\bibinfo {author} {\bibfnamefont {J.}~\bibnamefont
  {Preskill}},\ }\bibfield  {title} {\bibinfo {title} {Quantum {C}omputing in
  the {NISQ} era and beyond},\ }\href
  {https://doi.org/10.22331/q-2018-08-06-79} {\bibfield  {journal} {\bibinfo
  {journal} {{Quantum}}\ }\textbf {\bibinfo {volume} {2}},\ \bibinfo {pages}
  {79} (\bibinfo {year} {2018})}\BibitemShut {NoStop}%
\bibitem [{\citenamefont {Bharti}\ \emph {et~al.}(2022)\citenamefont {Bharti},
  \citenamefont {Cervera-Lierta}, \citenamefont {Kyaw}, \citenamefont {Haug},
  \citenamefont {Alperin-Lea}, \citenamefont {Anand}, \citenamefont {Degroote},
  \citenamefont {Heimonen}, \citenamefont {Kottmann}, \citenamefont {Menke},
  \citenamefont {Mok}, \citenamefont {Sim}, \citenamefont {Kwek},\ and\
  \citenamefont {Aspuru-Guzik}}]{GuzikRevModPhys2022}%
  \BibitemOpen
  \bibfield  {author} {\bibinfo {author} {\bibfnamefont {K.}~\bibnamefont
  {Bharti}}, \bibinfo {author} {\bibfnamefont {A.}~\bibnamefont
  {Cervera-Lierta}}, \bibinfo {author} {\bibfnamefont {T.~H.}\ \bibnamefont
  {Kyaw}}, \bibinfo {author} {\bibfnamefont {T.}~\bibnamefont {Haug}}, \bibinfo
  {author} {\bibfnamefont {S.}~\bibnamefont {Alperin-Lea}}, \bibinfo {author}
  {\bibfnamefont {A.}~\bibnamefont {Anand}}, \bibinfo {author} {\bibfnamefont
  {M.}~\bibnamefont {Degroote}}, \bibinfo {author} {\bibfnamefont
  {H.}~\bibnamefont {Heimonen}}, \bibinfo {author} {\bibfnamefont {J.~S.}\
  \bibnamefont {Kottmann}}, \bibinfo {author} {\bibfnamefont {T.}~\bibnamefont
  {Menke}}, \bibinfo {author} {\bibfnamefont {W.-K.}\ \bibnamefont {Mok}},
  \bibinfo {author} {\bibfnamefont {S.}~\bibnamefont {Sim}}, \bibinfo {author}
  {\bibfnamefont {L.-C.}\ \bibnamefont {Kwek}},\ and\ \bibinfo {author}
  {\bibfnamefont {A.}~\bibnamefont {Aspuru-Guzik}},\ }\bibfield  {title}
  {\bibinfo {title} {Noisy intermediate-scale quantum algorithms},\ }\href
  {https://doi.org/10.1103/RevModPhys.94.015004} {\bibfield  {journal}
  {\bibinfo  {journal} {Rev. Mod. Phys.}\ }\textbf {\bibinfo {volume} {94}},\
  \bibinfo {pages} {015004} (\bibinfo {year} {2022})}\BibitemShut {NoStop}%
\end{thebibliography}%

\end{document}